\documentclass{article}
\usepackage{epsfig}
\usepackage{float}
\usepackage{amsmath}

\newcommand{\be}{\begin{equation}}
\newcommand{\ee}{\end{equation}}

\newtheorem{theorem}{Theorem}[section]
\newtheorem{definition}[theorem]{Definition}
\newtheorem{proposition}[theorem]{Proposition}

\begin{document}

\title{Oscillations of networks: the role of soft nodes}

\author{ Jean-Guy CAPUTO$^{1}$ , Arnaud KNIPPEL$^{1}$ and Elie SIMO$^{1,2}$}

\maketitle

{\normalsize \noindent $^1$:Laboratoire de Math\'ematiques, INSA de Rouen,\\
B.P. 8, 76131 Mont-Saint-Aignan cedex, France\\
E-mail: caputo@insa-rouen.fr \\
$^2$: Physics Department, \\
Faculty of Sciences, University of Yaounde I,\\
B. P. 812 Yaounde - Cameroon}

\date{\ }

\begin{abstract}
To describe the flow of a miscible quantity on a network, we consider 
the graph wave equation where the standard continuous Laplacian 
is replaced by the graph Laplacian.
The structure of the graph influences strongly the dynamics which is naturally
described using the eigenvectors as a basis. 
Assuming the graph is forced and damped at specific nodes, 
we derive the amplitude equations. These reveal the importance of
a soft node where the eigenvector is zero. 
For example forcing the network at a resonant frequency 
reveals that damping can be ineffective if applied to such a soft node, leading
to a disastrous resonance and destruction of the network. 
We give sufficient conditions for the existence of soft nodes and
show that these exist for general graphs so that they
can be expected for complex physical networks and 
engineering networks like power grids.
\end{abstract}

\section{Introduction}

The flow of a scalar quantity in a network is an important
problem for fundamental science and engineering applications.
The latter include gas, water or power distribution networks.
Other examples are a simplified version of road traffic and the
flow of nerve impulse in the brain. The static aspect of the
problem has long been studied within the framework of operations
research, see for example \cite{minoux}. However in many cases the dynamic character is crucial.
Take for example the prediction of traffic jams in a given road 
network or the prediction of a flood in a river basin.

The basic equations describing the flow of a miscible quantity are
the well-known conservation laws of mathematical physics. 
These laws are building blocks for studying physical systems.
They are universal and can be found for example in mechanics and in 
electromagnetism.
Each conservation law has the form of a flux relation
\be\label{gen_cons_law}
u_t + \partial_x q = 0,
\ee
where the first term is the time derivative of a density $u$ and the second
one is the space derivative of the flux $q$
along the relevant coordinate.
The most important conserved quantities in mechanics are
the mass, momentum and energy.
In electromagnetism the current and voltage obey conservation
laws.

To describe the flow of a given quantity on a
network it is natural to generalize these conservation laws. 
For that one uses the graph representing the network. The derivative
is now the generalized gradient $\nabla$ or its transpose, the 
incidence matrix. 
To write this, it is important to orient the branches of the graph.
In the static regime, these equations are usually solved as an
optimization problem, see \cite{gas} for an example of a gas network.
The dynamic problem is solved numerically using ordinary differential
equation solvers. In general, it is difficult to analyze the problem.
There are however, classes of systems which can be analyzed. This 
happens when dissipation exists only at the nodes and not on the
branches. 
A typical example is a small power grid where
the power line dissipation can be neglected and where only power
input and output occur at given nodes. If the system is linear and possesses 
some symmetry, the equations can be reduced to a graph wave equation as named
by Friedman and Tillich \cite{friedman01}. There, the usual Laplacian
is replaced by its discrete analog, the graph Laplacian $ \nabla^T \nabla$.
The graph wave equation was studied by Maas 
\cite{maas} who considered graphs obtained by linking elementary
graphs. He established particular inequalities for the first 
non zero eigenvalue of the Laplacian of these graphs. 
The graph Laplacian was
also considered recently by Burioni et al for the thermodynamics of
the Hubbard model to describe static configurations of a Josephson junction
array \cite{burioni_jj}. Another problem considered by that group is
to use the graph as a controlled obstacle to obtain a desired reflection
of discrete nonlinear Schr\"{o}dinger solitons \cite{burioni_dnls}.

In this article we adopt a more applied point of view. We
recall how conservation laws lead to the graph wave equation. 
This is illustrated in three different
physical contexts: a network of inductances and capacities, its 
equivalent  mechanical analog represented by masses and springs and
an array of fluid ducts. From another point of view, the graph
wave equation can describe small oscillations of a network near
its functioning point. The key point is that all nodes have the
same inertia in the dynamics. Then the formulation using the graph wave 
equation is very useful because the graph Laplacian i.e. the 
spatial part of the equation is a symmetric
matrix so that its eigenvalues are real and its eigenvectors
are orthogonal. It is then natural to describe the dynamics of the
network by projecting it on a basis of the eigenvectors.
We consider that the network is fixed and is submitted to
forcing and damping on specific nodes and obtain the new amplitude 
equations. These reveal that when a component of an eigenvector for
a mode is zero the forcing or damping is ineffective. This concept
of a "soft node" is formalized. We then show what are the sufficient
conditions for general graphs to present such soft nodes. The numerical
analysis of a network forced close to resonance confirms the initial 
observation. When the network is damped on a soft node, the damping is 
ineffective. This can have catastrophic consequences because the amplitude of
the oscillations increases without bounds. Another notable effect 
occurs when a multiple eigenvalue corresponds to different eigenvectors.
Then depending how the system is excited we get a different response.
These are effects that can occur for a general
network and can have important practical consequences. \\
The article is organized as follows, section 2 presents the derivation of the
graph wave equation in different physical contexts. In section 3 we formalize
the notion of "soft node" and give sufficient conditions for a graph 
to have one. The amplitude equations are analyzed numerically in section 5.
There we force the simple graphs of section 3 at resonance and analyze their
response. We conclude in section 6.

\section{The model: graph wave equation}
\begin{figure}
\centerline{
\epsfig{file=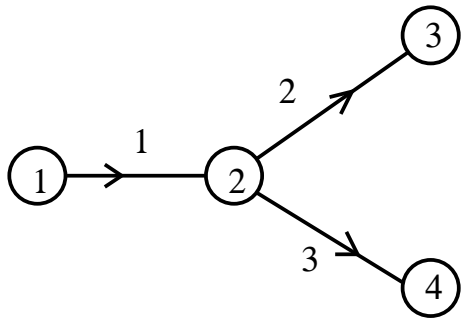,width=0.5\linewidth,angle=0}}
\caption{Schematic drawing of the complex graph oscillator. The arrows on the
branches are oriented arbitrarily.}
\label{f1}
\end{figure}

We now introduce the basic notions from graph
theory following the presentation of \cite{bls07}.
A graph ${\cal G}(N,B)$ is the association of a node set $N$ and
a branch set $B$ where a branch is an unordered pair of distinct
nodes. We assume that the nodes and branches are 
numbered $N=\{ 1,2,\dots n \}$
and $B=\{ 1,2,\dots m \}$ and that $n>1$ and $m>0$. The vertices
are oriented with an arbitrary but fixed orientation.
We consider for simplicity
only simple graphs which do not have multiple branches. In the following we
will refer to 
Fig. \ref{f1} shows such a graph with $n=4$ nodes
and $m=3$ branches.
The basic tool for expressing a flux is the so-called incidence matrix
$A(n,m)$ defined as
\begin{eqnarray}
\label{def_incidence}
A_{nb}  = -1 ~~  {\rm~ if ~branch}~ b~ {\rm starts~ from~ node~ }n, \\
A_{nb} = 1  ~~  {\rm~ if ~ branch}~ b~ {\rm finishes~ at~ node~ }n
,\\
A_{nb} = 0 {\rm~ otherwise}
\end{eqnarray}
In this section we present the conservation laws and derive the
graph wave equation using as example
the graph of Fig. \ref{f1}. This is done for pedagogical
reasons so the article is easier to read for non specialists. 
All these results can be easily generalized to any graph.
For the example shown in Fig. \ref{f1}, we have
$$A = \left ( \begin{matrix}
-1 & 0 & 0 \cr
1 & -1 & -1 \cr
0 & 1 & 0  \cr
0 & 0 & 1 
\end{matrix}  \right) $$
The transpose $A^T=\nabla$ is a discrete differential operator (
gradient of graph). To see this consider a function
$f: N \to R 
$. The vector $\nabla f$ has as $b$ component the difference of the values of
$f$ at the end points of the branch $b$ (with orientation). In the
example above, we have
\be\label{ex_nabla}
\nabla f = \left ( \begin{matrix}
-1 & 1 & 0 & 0 \cr
0 & -1 & 1 & 0 \cr
0 & -1 & 0 & 1 
\end{matrix}  \right)
\left ( \begin{matrix}
f_1 \cr
f_2 \cr
f_3 \cr
f_4 
\end{matrix}  \right)
= \left ( \begin{matrix}
f_2 - f_1 \cr
f_3 - f_2 \cr
f_4 - f_2 
\end{matrix}  \right),
\ee
which is the discrete gradient of $f$ associated to the graph.

\begin{figure}
\centerline{
\epsfig{file=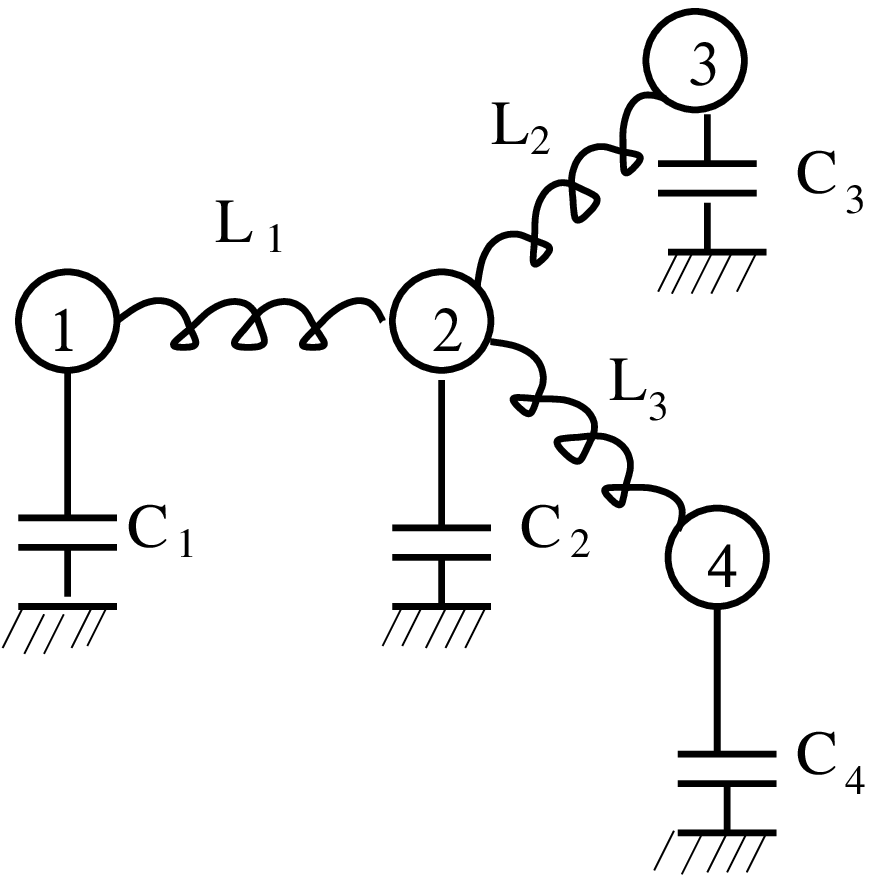,height=5 cm,width=6 cm,angle=0}
\epsfig{file=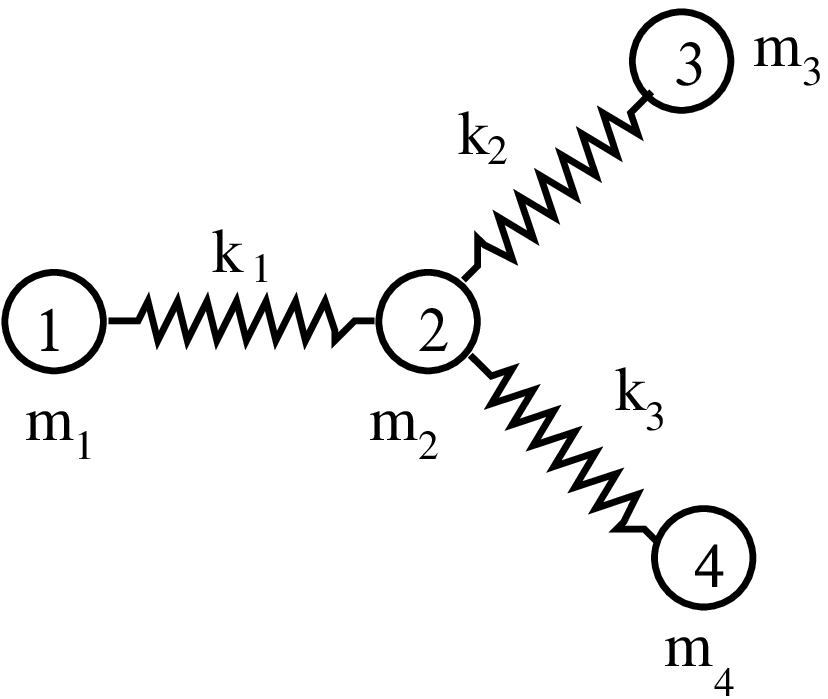,height=5 cm,width=6  cm,angle=0}
}
\caption{Two different physical networks corresponding to the graph shown in Fig. \ref{f1}.
The left panel shows a network of inductances and capacities and the right panel shows
the mechanical analog in terms of masses and springs}
\label{f1a}
\end{figure}

We now consider different networks where this formalism applies. We will
first write conservation laws using the $\nabla$ operator or its transpose.
From there we establish the relevant wave equation. These derivations
are shown for the example of Fig. \ref{f1} for pedagogical reasons, they
can be generalized to any graph.
For the specific inductance-capacity 
electrical network shown in the left panel
of Fig. \ref{f1a}, the equations of motion in terms
of the (node) voltages and (branch) currents are
the conservation of current and voltage
\begin{eqnarray}\label{iv}
C v_t - \nabla^T i = s,\\
L i_t - \nabla v = 0 \nonumber,
\end{eqnarray}
where 
$$C= \left ( \begin{matrix}
C_1 & 0 & 0 & 0 \cr
0 & C_2 & 0 & 0 \cr
0 & 0 & C_3 & 0 \cr
0 & 0 & 0 & C_4 
\end{matrix}  \right),~~~L = \left ( \begin{matrix}
L_1 & 0 & 0  \cr
0 & L_2 & 0  \cr
0 & 0 & L_3 
\end{matrix}  \right) ~, $$
are respectively the diagonal matrix of capacities and the diagonal 
matrix of inductances and $s$ represents the currents applied to each node.  
(similar to boundary conditions in the continuum case). Note that the
equations (\ref{iv}) also describe in the linear limit, 
shallow water waves in a network of canals and
fluid flow in a pipe network. In the first situation
$v$ corresponds to a surface elevation and $i$ to a flux. For the second example
$v$ is the pressure and $i$ the flux.
From the two equations (\ref{iv}) one obtains 
the generalized wave equation. For this take the derivative of the 1st 
equation and substitute the second to obtain the node wave
equation
\be\label{lc_wave}
Cv_{tt} - \nabla^T L^{-1} \nabla v = s_t,\ee
where 
\be\label{gen_laplacian} \nabla^T L^{-1} \nabla = \left ( \begin{matrix}
L_1^{-1} & -L_1^{-1} & 0 & 0 \cr
-L_1^{-1} & L_1^{-1} + L_2^{-1} +L_3^{-1} & - L_2^{-1} & - L_3^{-1} \cr
0 & -L_2^{-1} & L_2^{-1} & 0 \cr
0 & -L_3^{-1} & 0 & L_3^{-1}
\end{matrix}  \right)~. \ee
There is a corresponding branch wave equation for the 
currents that involves the
link Laplacian in the terminology of \cite{friedman01}. Taking the time
derivative of the second equation and substituting into the first we get
\be\label{lc_wave2}
L i_{tt} - \nabla C^{-1} \nabla^T i = \nabla C^{-1} s.\ee
In the rest of the article we will only consider the node graph wave
of the type (\ref{lc_wave}).

A similar graph wave equation arises when describing the other physical system
shown in the right panel of Fig. \ref{f1a}, the collection of four masses
$m_i,~i=1-4$ connected by springs of stiffnesses $k_i,~i=1-3$. Here the
variables are the displacements $y_i, ~i=1-4$ of each mass. The equations
of motion are
$$
\left\{
\begin{array}{llll}
m_1\ddot{y_1}=k_1\left(y_2-y_1\right) \\
m_2\ddot{y_2}=k_1\left(y_1-y_2\right) + k_2\left(y_3-y_2\right)
+k_3\left(y_4-y_2\right) \\
m_3\ddot{y_3}=k_2\left(y_2-y_3\right) \\
m_4\ddot{y_4}=k_3\left(y_2-y_4\right).\\
\end{array}
\right.
$$
Notice the correspondence capacities / masses and inverse 
inductances / stiffnesses. The matrix on the right hand side is symmetric
just like the matrix in the second term on the left hand side of 
(\ref{lc_wave}).

\begin{figure} [H]
\centerline{\epsfig{file=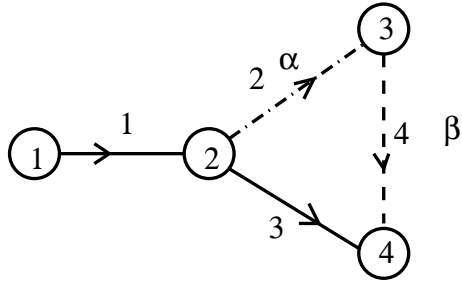,width=0.5\linewidth,angle=0}}
\caption{The specific network that will be studied throughout the article.}
\label{f1b}
\end{figure}

In the following we will consider the graph shown in Fig. \ref{f1b} which
includes an additional branch between nodes 3 and 4. We assume that
the masses are equal. This is important because it preserves the symmetry
of the operator. For simplicity we choose $k_1=k_3=k$, $k_2=\alpha^2 k$ and
$k_4 = \beta^2 k$.
Then we normalize times by the natural frequency 
\begin{equation}
\omega=\sqrt{\frac{k}{m}},~~
t\prime=\omega t .
\end{equation}
Omitting the primes, the equations can be written in matrix form 
as
\begin{equation}
\left(
\begin{array}{c}
\ddot{y_1} \\
\ddot{y_2} \\
\ddot{y_3}\\
\ddot{y_4} \\
\end{array}
\right) =
\left(
\begin{array}{cccc}
-1 & 1 & 0 & 0 \\
1 & -2-\alpha & \alpha & 1 \\
0 & \alpha & -\alpha-\beta & \beta\\
0 & 1 &  \beta & -1-\beta \\
\end{array}
\right) 
\left(
\begin{array}{c}
y_1 \\
y_2 \\
y_3\\
y_4 \\
\end{array}
\right) ,
\label{cs1}
\end{equation}
which we will write formally as 
\be\label{lin_evo} Y_{tt} = G Y,\ee
where $G$ is the 
graph Laplacian \cite{bls07}, the equivalent of (\ref{gen_laplacian}).

Another way of defining the graph Laplacian $G$ is to consider
the adjacency matrix  $M=(m_{ij})$ where 
$m_{ij}$ is the weight of the branch connecting node $i$ to node $j$.
The weight is zero if the branch is absent.
The degree matrix $D=(d_{i})$ is the diagonal matrix such that
$$d_i = \sum_{j\neq i} m_{ij}.$$
We then have 
\be\label{def_glap}
G = \nabla^T \nabla = -D + M .\ee

In the rest of the article we keep the masses the same. 
Equation (\ref{cs1}) in the absence of $\alpha$ and $\beta$ is
a finite difference discretisation of the 1D continuous Laplacian. 
This connects the model with a continuous wave equation.
When $\alpha >0$ and $\beta>0$ we have a generalization of this finite
difference description. In a way we have a discrete version of 
coupled wave equations. 
From another point of view the graph wave equation above describes
small oscillations of a general (nonlinear) system around it's operating
state. It is a very general model.

The linear evolution problem (\ref{lin_evo})
gives rise to periodic solutions $Y(t) =Z \exp ({i \omega t}) $ where
the $Z$ verify the spectral problem
\be\label{spec_pb}
G Z = -\omega^2 Z .\ee
The matrix $G$ is symmetric.
If we had assumed different masses
on the nodes, we would have lost this property. 
For electrical networks this corresponds to the same capacity in
equation (\ref{lc_wave}). This property is important because symmetric
matrices have real eigenvalues and we can always build a basis
of orthogonal eigenvectors. These eigenvalues
are stable with small symmetric perturbations of the network, like for example
adding a link. In this sense the eigenmodes are characteristic of the
network.
The eigenvectors provide a basis of $R^n$
which is adapted to describe the evolution of $Y$ on the graph.
Specifically we arrange the eigenvalues $\lambda_i= -\omega_i^2$ as
\be\label{eigenvalues}
|\lambda_1 |= 0 \le |\lambda_2| \le \dots \le |\lambda_n|.\ee
We label the associated eigenvectors $v_1,v_2,\dots,v_n$. These verify
\be\label{eigenvectors}
G v_i = -\omega_i^2 v_i .\ee
and are orthogonal with respect to the standard scalar product. We then
normalize them so $< v_i ,v_j> = \delta_{i,j}$ where $\delta_{i,j}$ is
the Kronecker symbol.
Notice that $G$ is a singular matrix since the sum of its lines
(resp. columns) gives a 0 line (resp. column). Therefore $G$ will
have a zero eigenvalue which corresponds to the Goldstone mode.
The solutions of the linear system $G Y = S$ are given up to a constant.

It is natural to write the equation of motion (\ref{lin_evo}) in terms of the amplitudes of
the normalized eigenvectors $v_i, ~~Y = \sum_i a_i v_i$. We obtain the standard result
\be\label{evo_normalmode}
{\ddot a_i} + \omega_i^2 a_i = 0,\ee
so that the normal modes do not exchange energy. The problem is then Hamiltonian with
\be\label{ham_normalmode} H = {1\over 2}\sum_i {\dot a_i}^2 + \omega_i^2 {a_i}^2.\ee
When a perturbation acts on the system the equations need to be modified. This 
is the object of the next section.

\section{Forcing the network : amplitude equations}

For a general (nonlinear) evolution problem of the form
\be\label{nlin_evo}
Y_{tt} = G Y + N(Y),\ee
it is natural to expand $Y$ using the eigenvectors as
\be\label{expansion}
Y(t) = a_1(t)v_1 +a_2(t)v_2 + \dots + a_n(t) v_n . \ee
Inserting (\ref{expansion}) into (\ref{nlin_evo}) and projecting
on each mode $v_i$ we get the system of coupled equations
\begin{eqnarray}\label{evo_coeff}
{a_1}_{tt} + \omega_1^2 a_1 & =&  < N(Y) | v_1>, \\
{a_2}_{tt} + \omega_2^2 a_2 & =&  < N(Y) | v_2>, \\
 \dots  \\
{a_n}_{tt} + \omega_n^2 a_n & =&  < N(Y) | v_n> .
\end{eqnarray}
We expect this decomposition to be more adapted to describe the
dynamics of $Y$ on the graph. In particular it should explain
some of the unexpected couplings that are observed between the modes.

Let us now assume that the network is forced at some node $n_f$
and damped at some node $n_d$.
The motion can be represented as
\be\label{forced_graph}
Y_{tt} = G Y + D (-d Y_t) + F f {\bf 1} ,\ee
where the $(n,n)$ matrices $D,~F $ are everywhere 0 except for 
$D(n_d,n_d)=1 ,~F(n_f,n_f)=1$ and where $d$ is a damping 
coefficient and $f$ a forcing coefficient.
The vector ${\bf 1}$ is ${\bf 1} = (1,1,\dots, 1)^T$.
Assuming the linear combination for $Y$ (\ref{expansion}) and projecting
we get 
\be\label{ajt}
\ddot a_j = - \omega_j^2 a_j - d \sum_{k=1}^n <D v_k |v_j> 
{\dot a_k}
+ f <F {\bf 1} |v_j>.\ee
In terms of components we obtain
\be\label{cajt}
\ddot a_j = - \omega_j^2 a_j - d v_j^{n_d} \sum_{k=1}^n v_k^{n_d}
{\dot a_k}
+ f v_j^{n_f},~~j=1,n\ee
where 
$v_j^{n_d},~v_j^{n_f}$ are respectively the $n_d,~n_f$ components of 
the normal vector $v_j$.
From these equations one can see that exciting one
node will cause disturbances to propagate all through the network
in a precise way. The forcing (resp. the damping) will act on mode $j$ only 
if the $v_j^{n_f} \neq 0$ (resp. $v_j^{n_d} \neq 0$). 
This important fact leads to the introduction of the concept
of "soft node" which we formalize in the next section.

\section{Soft nodes}

We introduce the following definitions of 
a soft node for an eigenmode $j$ of the graph Laplacian $G$.
\begin{definition}[{\rm Soft node }]
\label{def1} 
A node $s$ of a graph is a soft node for an eigenvalue $\lambda$
if there exists an eigenvector $x$ for this eigenvalue such that $x_s=0$.
\end{definition}
Another definition is the one of absolute soft node.
\begin{definition}[{\rm Absolute soft node }]
\label{def2} 
A node $s$ of a graph is an absolute soft node for an eigenvalue $\lambda$
if every eigenvector $x$ for this eigenvalue is such that 
$x_s=0$.
\end{definition}
Note that for single eigenvalues, a soft node is an absolute soft node.

Then we have the following property
due to the relation (\ref{def_glap})
\be\label{sn}
\sum_{i\in\Gamma(s)} m_{is} x_i=(deg(s)+\lambda)x_s,\ee
where $deg(s)=\sum_{i\in\Gamma(s)} m_{is} $ where $\Gamma(s)$
is the set of nodes neighbors to $s$.
From this we deduce the following
\begin{proposition}
\label{lem1}
A soft node $s$ of a graph is such that 
$\sum_{i\in\Gamma(s)} m_{is} x_i=0$.
\end{proposition}





As illustration consider the graph shown in Fig. \ref{f1b} with $\beta=0$
and where $\alpha$ is a parameter. This graph is then a tree.
The eigenvalues and eigenvectors can be computed analytically, they
are given in Table 1 in terms of $\alpha$ and
$$\delta = \sqrt{4\alpha^2-4\alpha+9}.$$
\begin{table} \label{tab2}
\begin{tabular}
{|l | c | c | c | r |}
  \hline
index $i $           &  1    &  2 &  3   &   4 \\ \hline 
$\lambda_i= -\omega_i^2$   &  0    & -1 & ${1\over 2}(\delta-2\alpha-3)$ &$-{1\over 2}(\delta +2\alpha +3)$\\ 
                           &       &    &                                &                                 \\\hline
                           &  1    & 1   &  1                            &  1                             \\ 
$v_i$                      &  1    & 0   & ${1\over 2}(\delta-2\alpha-1)$ & $ -{1\over 2}(\delta + 2\alpha +1)$ \\ 
                           &  1    & 0   & $-{1\over 2}(\delta-2\alpha +3)$& $ {1\over 2}(\delta + 2\alpha -3 )$\\ 
                           &  1    & -1  &  1                            &   1                            \\ \hline
 \end{tabular}
\caption{Eigenmodes $(\lambda_i, v_i)$ for the tree.}
\end{table}

\begin{figure} [H]
\centerline{\epsfig{file=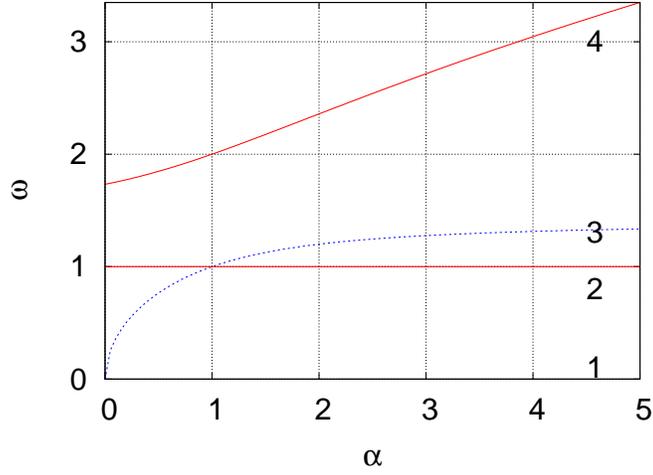,width=0.8\linewidth,angle=0}}
\caption{ Plot of the eigenfrequencies $\omega_i,~i=1-4$ as a function of 
$\alpha$ for the graph shown in Fig. 3 with $\beta=0$ i.e. a tree.}
\label{f1c}
\end{figure}

The frequencies $\omega_i=\sqrt{-\lambda_i}$ are plotted as a function of 
the parameter $\alpha$ in Fig. \ref{f1c}. 
Note how the eigenvalue $\omega_2=1$ is independent of $\alpha$.
This is because the graph has a swivel at node $3$ as we will explain below.
Table 1 shows that $s=2$ and
$s=3$ are absolute soft nodes for $\lambda=-1$ and $\alpha\neq 1$. When
$\alpha=1$ the eigenvalue $\lambda=-1$ is degenerate and two eigenvectors
are
$(1,0,0,-1)$ and $(1,0,-2,1)$. The node $s=2$ remains an absolute
soft node while the node $s=3$ is soft.

\begin{figure} [H]
\centerline{
\epsfig{file=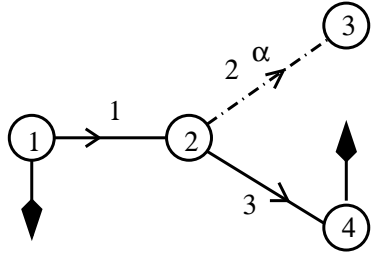,width=0.4\linewidth,angle=0}}
\caption{Schematic representation of
the constant mode $v_2$, corresponding to  $\omega_2=1$}.
\label{f3a}
\end{figure}

The mode $v_2$ which is independent of $\alpha$ is shown schematically
in Fig. \ref{f3a} where we have plotted the magnitude and sign 
of the coordinate using vertical arrows. The arrows are opposite and equal
for $v_{2}^1$ and $v_{2}^4$ because the nodes 1 and 4 play a symmetric
role i.e. the graph is invariant by the automorphism transforming node
1 to node 4 \cite{crs01}. 
This representation of the eigenmodes is intuitive and gives a direct
information on the coupling of the node to the eigenmode. We will
use this representation throughout the article. 

\begin{figure} [H]
\centerline{
\epsfig{file=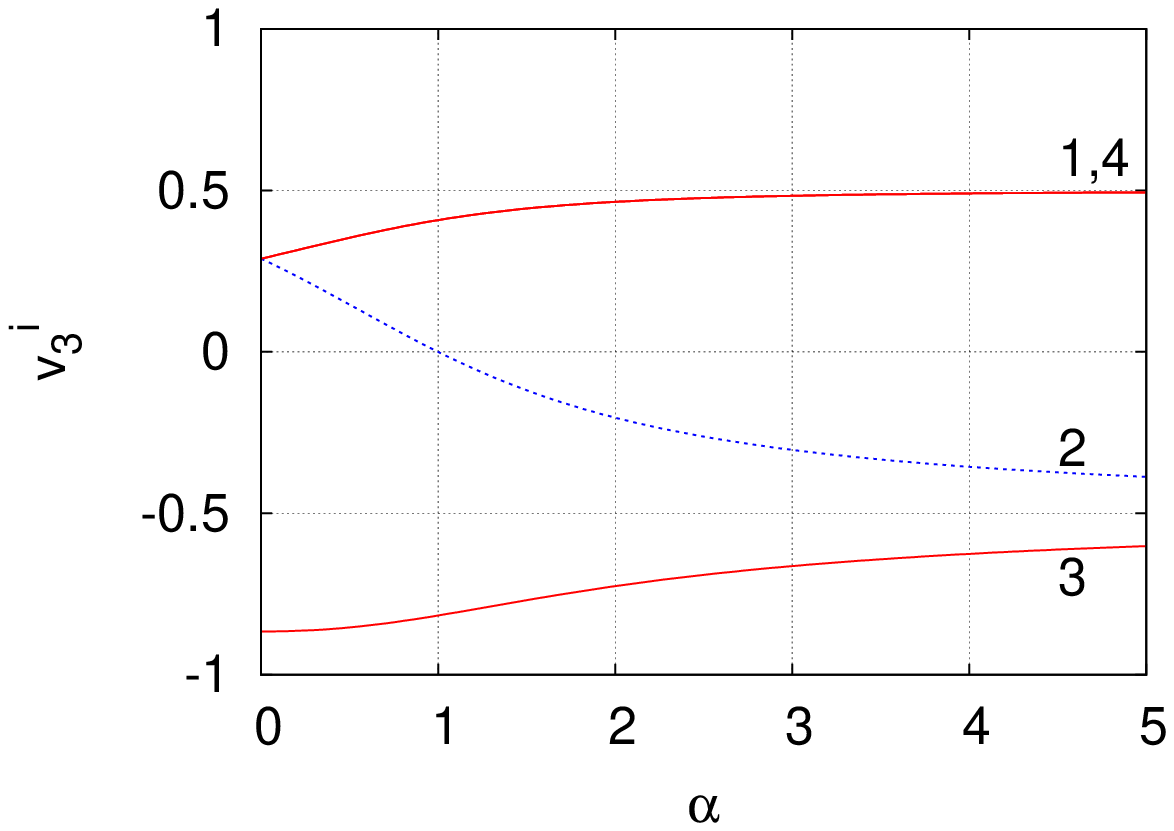,width=0.5\linewidth,angle=0}
\epsfig{file=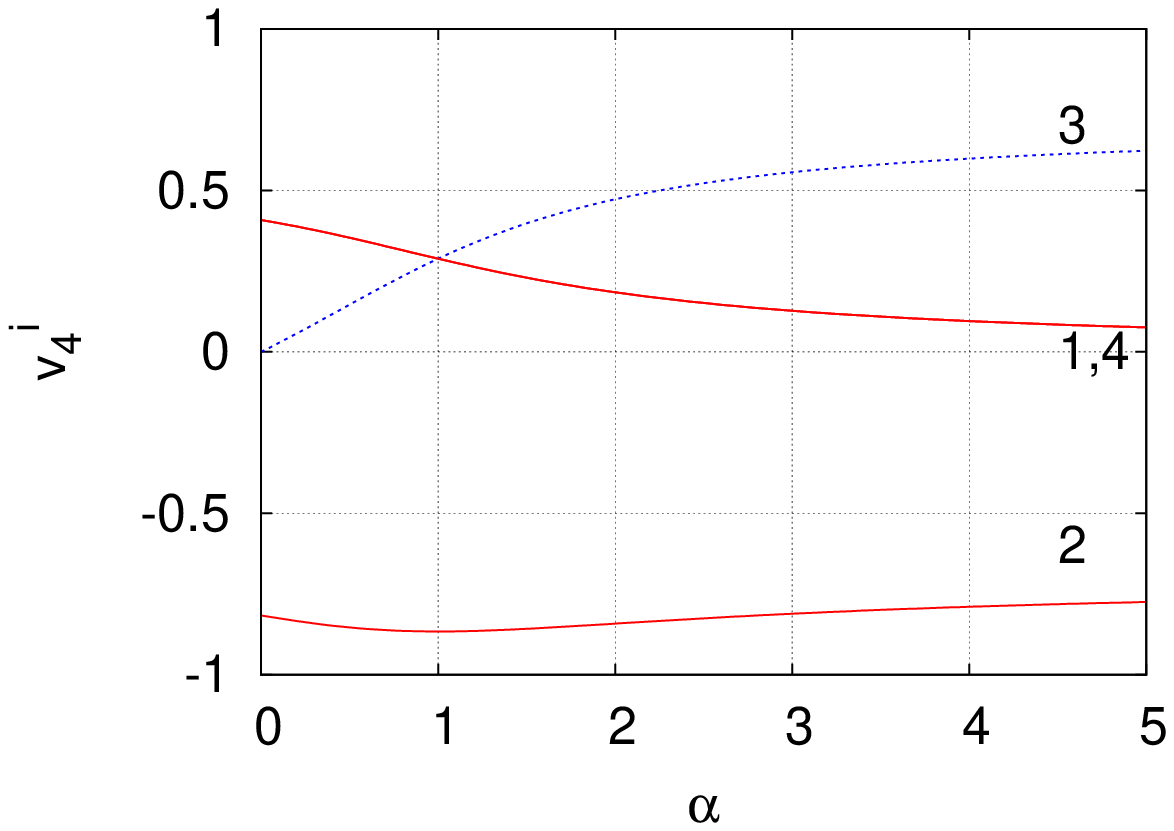,width=0.5\linewidth,angle=0}
}
\caption{Plot of the components of the non trivial normalized eigenvectors $v_3$
(left panel) and $v_4$ (right panel)
as a function of $\alpha$. {\bf soft node for $\alpha=1$ and $v_3$} }
\label{f2}
\end{figure}

Let us now consider the other two modes $v_3$ and $v_4$ which are
dependant on $\alpha$.  Fig. \ref{f2} shows the plots of the
components $v_3^i$ and $v_4^i$. For $\alpha =1$ we see that
the node $s=2$ is soft for the mode $v_3$. We will see below that
this feature enables to set the functioning point of the network.
For $\alpha >5$ 
the magnitude of $v_4^1$ and $v_4^4$
is much smaller than the magnitude of $v_4^2$ and $v_4^3$ so that the
nodes 4 and 1 are close to soft. Notice how nodes 2 and 3 are soft
for $v_2$ while nodes 1 and 4 are soft for $v_4$ when $\alpha >>1$.
Anticipating on the dynamics, one sees that
for such large values of $\alpha$,
when the graph is forced at a 
frequency $\omega\approx 1$ the links 1 and 3 will be active and the
nodes 2 and 3 will be soft. On the contrary if the graph is forced at
a high frequency $\omega\approx \omega_4$ only link 2 will be active 
and nodes 1 and 4 will be soft.

\begin{figure} [H]
\centerline{
\epsfig{file=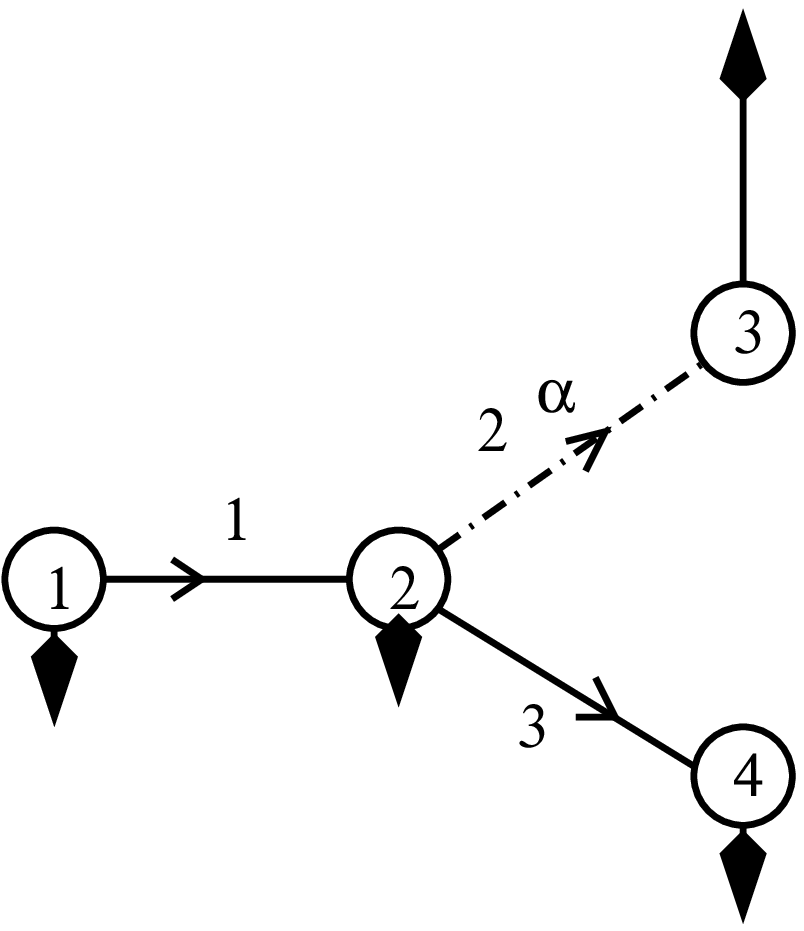,width=0.3\linewidth,angle=0}
\hskip 0.3 cm
\epsfig{file=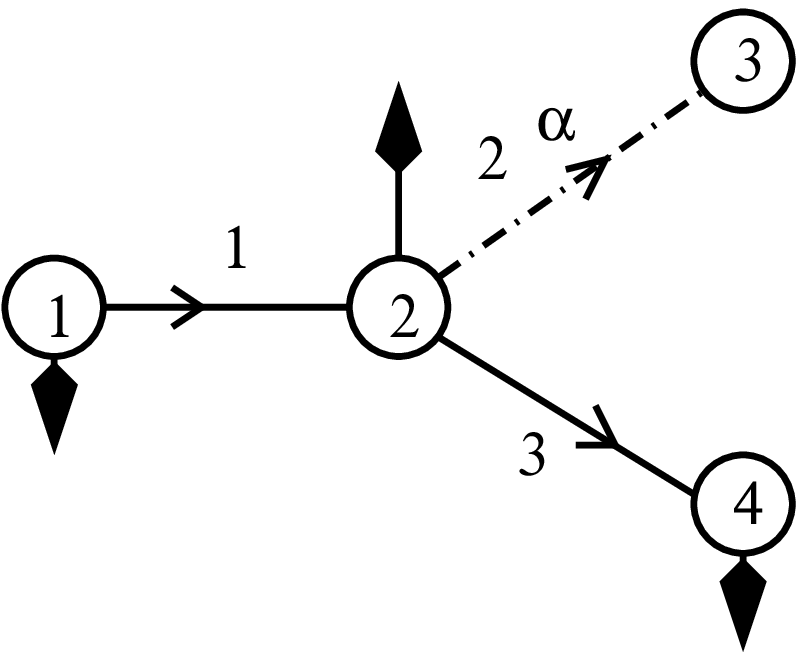,width=0.3\linewidth,angle=0}
\hskip 0.3 cm
}
\centerline{
\epsfig{file=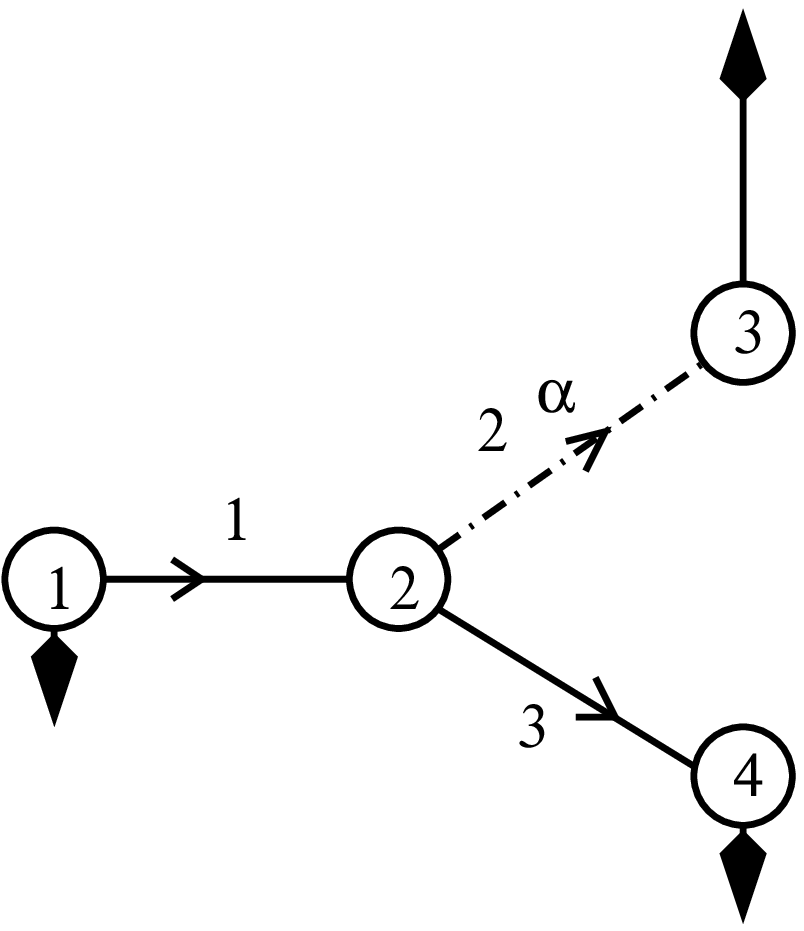,width=0.3\linewidth,angle=0}
\hskip 0.3 cm
\epsfig{file=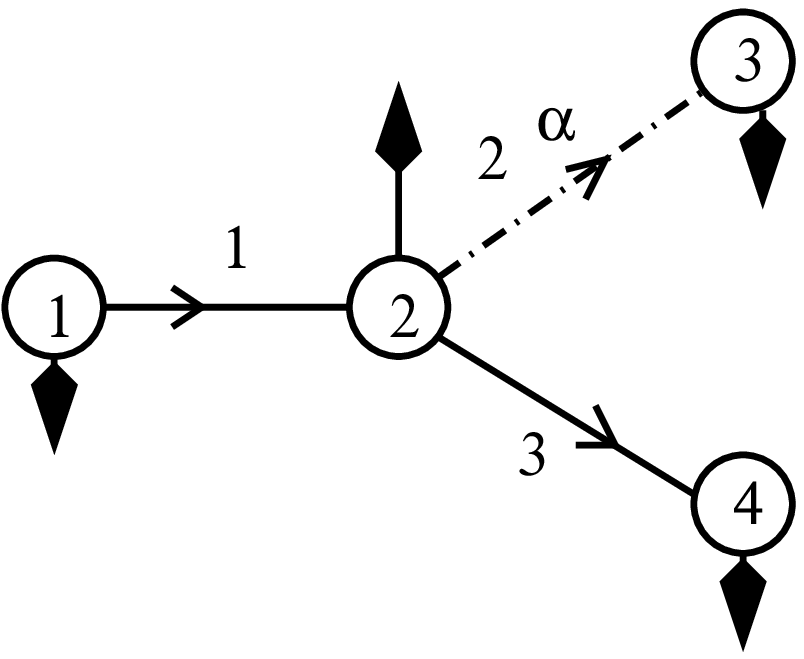,width=0.3\linewidth,angle=0}
\hskip 0.3 cm
}
\centerline{
\epsfig{file=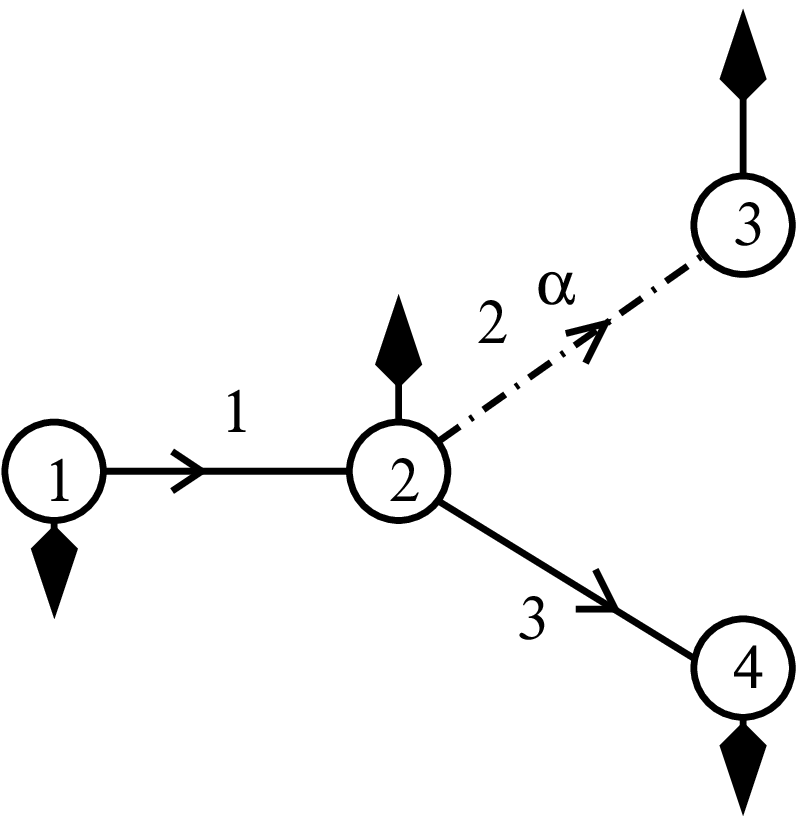,width=0.3\linewidth,angle=0}
\hskip 0.3 cm
\epsfig{file=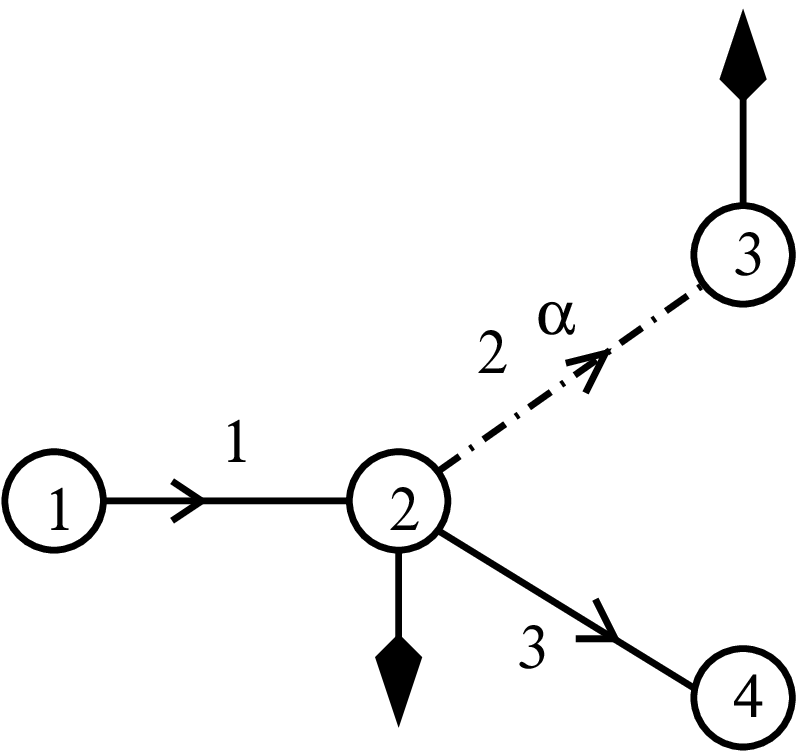,width=0.3\linewidth,angle=0}
\hskip 0.3 cm
}
\caption{Schematic representation of the non trivial 
eigenvectors $v_3,v_4$ shown respectively on the left 
and the right panels.  
The top panel is for $\alpha=0.1$. It shows the modes 
$v_3$ corresponding to $\omega_3=0.36$ (left) and  
$v_4$ corresponding to $\omega_4=1.75$ (right). The
middle panel is for $\alpha=0.9$ and shows the
modes $v_3$ corresponding to $\omega_3=0.964$ (left) and
$v_4$ corresponding to $\omega_4= 1.9671$ (right).
The bottom panel corresponds to $\alpha=5$. It
shows the modes $v_3$ corresponding to $\omega_3=1.335$ (left) and
$v_4$ (right) corresponding to $\omega_4=3.35$.}
\label{f3}
\end{figure}
All this information is summarized in 
Fig. \ref{f3} which show a schematic representation 
of the non trivial
eigenvectors for $\alpha=0.1,~0.9$ and 5. One can easily identify the
soft nodes. 
As expected by the
graph automorphism the components on nodes 1 and 4 of the modes
are equal.


Let us return to the $\alpha$-independent eigenvalue $\lambda_2=-1$
and its eigenvector $v_2$ with soft nodes 2 and 4. These properties 
exist for general graph configurations of the swivel type which we define
in the next section.

\subsection{The swivel : a graph with a soft node}

Recall that a leaf is a node connected to only one other node.

\begin{definition}[{\rm A swivel }]
\label{def3}
Given a graph, a swivel (or swivel node) of order $k$ is a node connected
to $k$ leaves with all the same coupling $\alpha$.
\end{definition}

We have the following property.
\begin{proposition}
\label{lem2}
For a graph with a swivel of order 2 and coupling $\alpha$, 
the Laplacian $G$ has an eigenvalue $\lambda=-\alpha$. The swivel
node and the other nodes except the leaves are soft for this eigenvalue.
\end{proposition}

\begin{figure} [H]
\centerline{
\epsfig{file=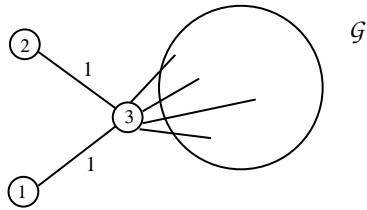,width=0.4\linewidth,angle=0}}
\caption{A graph such that two leaves are connected to
a swivel (with the same coupling).}
\label{fgen}
\end{figure}

{\bf Proof }\\
We consider the situation shown in Fig. \ref{fgen}
where the outer nodes 1 and 2 are connected to a common node 3
which has $p$ connections to the rest ${\cal G}'$ of the graph.
Assuming $\lambda$ to be an eigenvalue of $G$ we write the matrix $G - \lambda I$ 
\be\label{caract_pol}
G - \lambda I = 
\left (
\begin{array}{ccccccccc}
-\alpha-\lambda & 0 & \alpha & 0 & \dots & \dots & \dots &\dots &0 \\
0 & -\alpha-\lambda  & \alpha & 0 & \dots & \dots & \dots &\dots &0 \\
\alpha & \alpha  & -p-2\alpha-\lambda & m_{34} & \dots &  &  & & m_{3n} \\
0 & 0  & m_{34} &  &  & &  & & \\
| & |  & | &  &  & &  & & \\
| & |  & | &  &  & A' &  & & \\
| & |  & | &  &  & &  & & \\
| & |  & | &  &  &  &  & & \\
0 & 0  & m_{3n} &  &  & &  & & 
\end{array}
\right ) ,
\ee
where the submatrix $A'$ corresponds to the subgraph ${\cal G'}$
and 
$$p = \sum_{k=4}^n m_{3k}.$$
It is easy to see that if $\lambda=-\alpha$ then the vector $v_1$ of coordinates
$(x_1,x_2,\dots,x_n)$ where $x_1+x_2=0$ and $x_i=0,i=3,\dots n$ 
is an eigenvector because $(G+I)v_1 =0$. 
Q.E.D.
 

We now generalize this result to swivels of order $k>1$.
The matrix $G - \lambda I$ now has $k$ lines similar to the third line of 
(\ref{caract_pol}). 
Following a similar argument as above it can be shown that

\begin{proposition}
\label{lem3}
If $k$ leaves of a graph are connected to a common swivel $s$ with
the same coupling $\alpha$ then $-\alpha$ is an eigenvalue of multiplicity
$k-1$. Except maybe the leaves linked to the swivel, all the nodes
are soft for the eigenvalue $-\alpha$.
The eigenvectors for this eigenvalue will then be such that
$$\sum_{i \in \Gamma(s)} x_i = 0,$$
\end{proposition}
where $\Gamma(s)$ is the set of the nodes adjacent (neighbors) to node $s$.

\subsection{The closed swivel : another graph with a soft node}

Let us now consider that the branches of an order 2 swivel are connected. Then
we talk about a graph with a closed swivel.
\begin{definition}[{\rm A closed swivel }]
\label{def4}
A graph with a closed swivel is constructed by connecting
the two leaves of a swivel of order 2 and coupling $\alpha$ by 
a branch of coupling $\beta$.
\end{definition}
We have the following property for the eigenvalues and associated eigenmode
of coordinates $x_i$.
\begin{proposition}
\label{lem4}
Assume a graph with a closed swivel of coupling $\beta$. 
Then the Laplacian has eigenvalue $-1-2\beta$. For this mode, the swivel is
a soft node $x_3=0$, $x_1=-x_2$ and $x_k=0,~k=4\dots n$.
\end{proposition}
Note that for $\beta=0$ we recover the result of the previous subsection.

{\bf Proof }\\
Following the labeling of the graph presented above and summarized in
Fig. \ref{fgen}, we assume that the nodes 1 and 2 are connected by an 
additional link of coupling $\beta$.
The matrix $G - \lambda I$ is then
\be\label{caract_pol2}
G - \lambda I =
\left (
\begin{array}{ccccccccc}
-\alpha-\lambda-\beta & \beta & \alpha & 0 & \dots & \dots & \dots &\dots &0 \\
\beta & -\alpha-\beta-\lambda  & \alpha & 0 & \dots & \dots & \dots &\dots &0 \\
\alpha & \alpha  & -p-2\alpha-\lambda & m_{34} & \dots &  &  & & m_{3n} \\
0 & 0  & m_{34} &  &  & &  & & \\
| & |  & | &  &  & &  & & \\
| & |  & | &  &  & A' &  & & \\
| & |  & | &  &  & &  & & \\
| & |  & | &  &  &  &  & & \\
0 & 0  & m_{3n} &  &  & &  & &
\end{array}
\right ) ,
\ee
Let us show that $\lambda=-\alpha-2\beta$ is an eigenvalue. The form 
of the equations leads 
to assume that $x_i=0, i \ge 4$. Then the equations for $x_1,x_2,x_3$ are
\begin{eqnarray}
\beta x_1 + \beta x_2 +\alpha x_3=0,\\
\beta x_1 + \beta x_2 +\alpha x_3=0,\\
\alpha x_1 +\alpha x_2 +(-p-2-\lambda)x_3=0.
\end{eqnarray}
which have as only solutions $x_1+x_2=0$ and $x_3=0$. This proves that
$-\alpha-2\beta$ is an eigenvalue of $G$, that the swivel is a soft node $x_3=0$
and that the eigenvector is such that $x_1+x_2=0$ and $x_i=0, i\ge 4$.
Q.E.D.

\begin{figure} [H]
\centerline{\epsfig{file=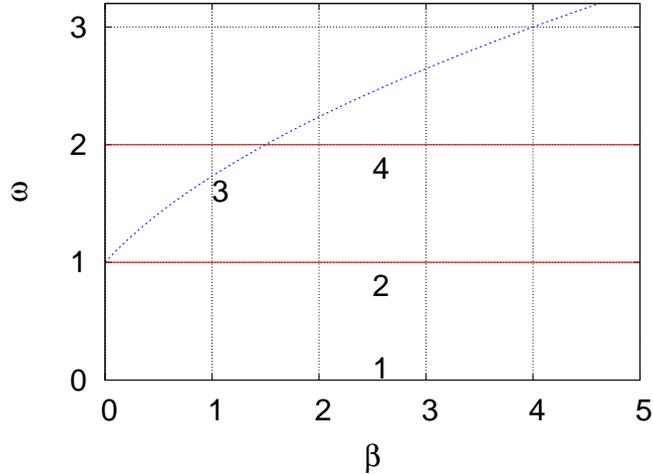,width=0.8\linewidth,angle=0}}
\caption{ Plot of the eigenfrequencies $\omega_i,~i=1-4$ as a function of
$\beta$ for the graph shown in Fig. 3 with $\alpha=1$. 
}
\label{f1e}
\end{figure}

\begin{figure} [H]
\centerline{
\epsfig{file=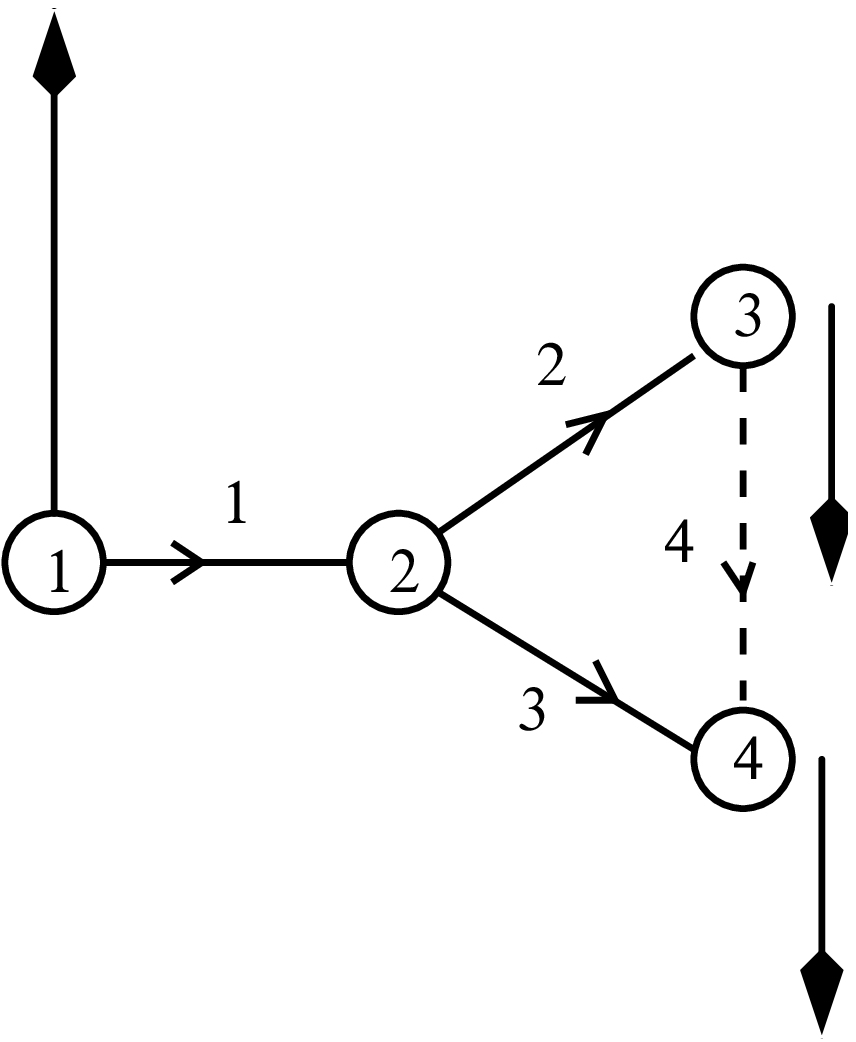,width=0.2\linewidth,angle=0}
\hskip 0.3 cm
\epsfig{file=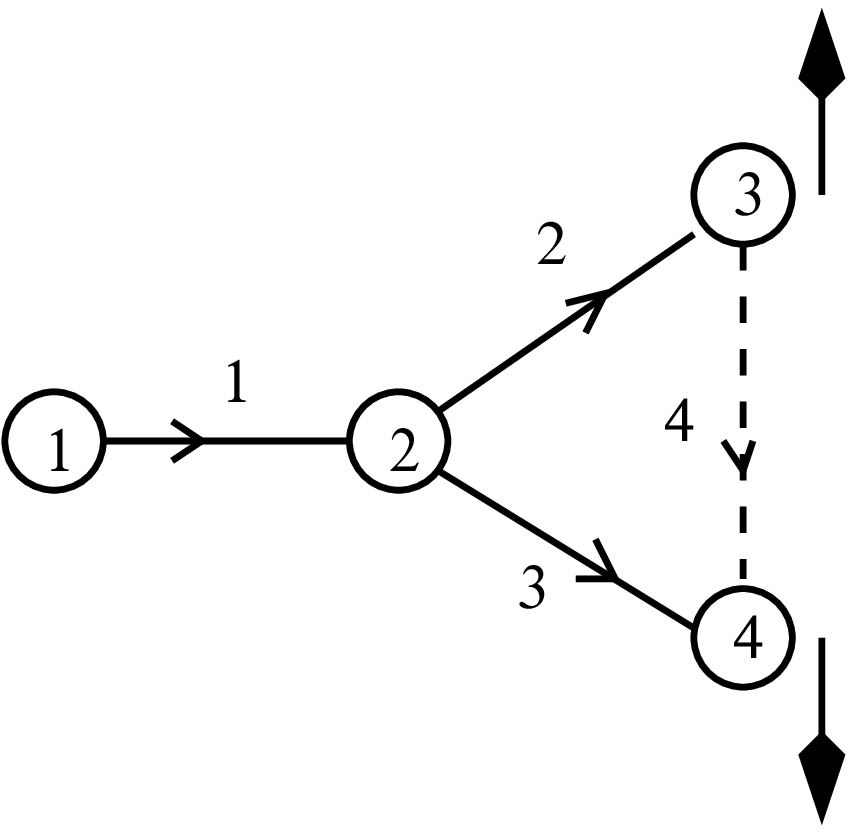,width=0.2\linewidth,angle=0}
\hskip 0.3 cm
\epsfig{file=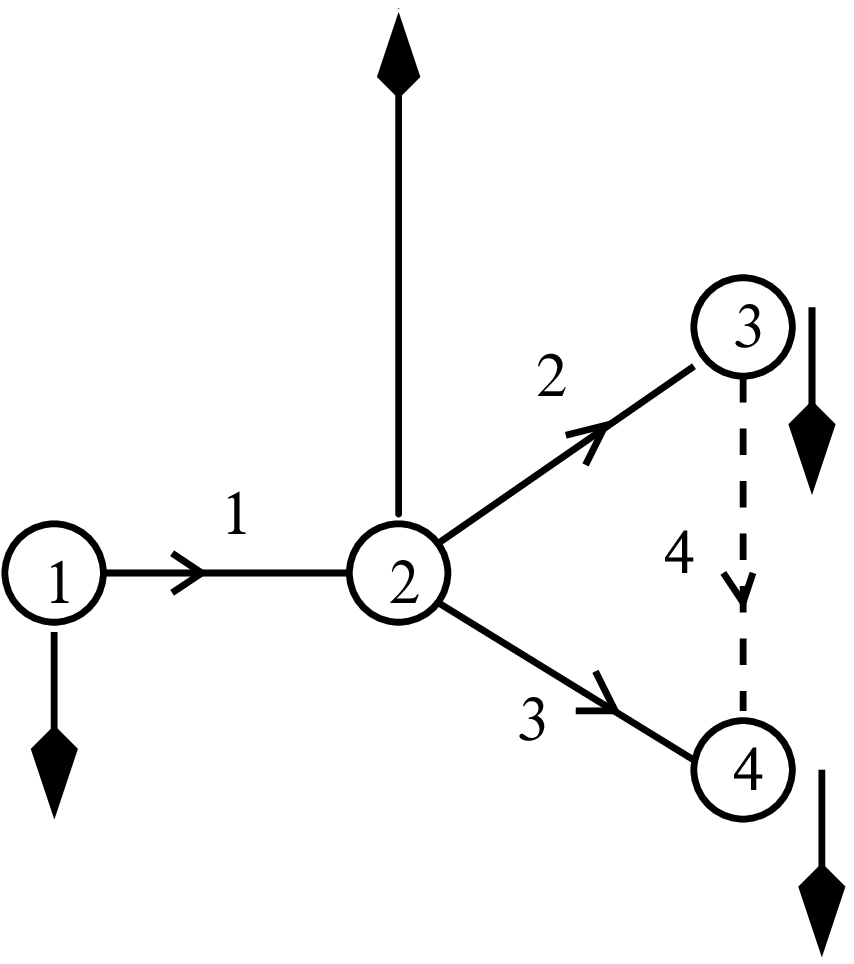,width=0.2\linewidth,angle=0}
\hskip 0.3 cm
}
\caption{Schematic representation of the eigenvectors
$v_2$, $v_3$ and $v_4$ , from left to right}
\label{f1f}
\end{figure}

A simple closed swivel is the graph shown in (\ref{f1b}) where we
assume $\alpha=1$ and vary $\beta$. The eigenmodes can be computed
analytically. They are presented in table 2.
\begin{table} \label{tab3}
\begin{tabular}
{|l | c | c | c | r |}
  \hline
index $i $           &  1    &  2 &  3   &   4 \\ \hline
$\lambda_i= -\omega_i^2$   &  0    & -1 & $ -1 -2 \beta $ & $-4$ \\
                           &       &     &                                &                                 \\\hline
                           &  1    & 1     & 0  &  1    \\
$v_i$                      &  1    & 0     & 0  &  -3 \\ 
                           &  1    & -1/2  & 1  &  1\\   
                           &  1    & -1/2  & -1 &   1   \\ \hline
 \end{tabular}
\caption{Eigenmodes $(\lambda_i, v_i)$ for the graph with $\alpha=1$
and a varying $\beta$.}
\end{table}
The eigenfrequencies $\omega_i=\sqrt{-\lambda_i}$ are shown in Fig. \ref{f1e} as
a function of $\beta$. As expected we find the eigenvalues $-1$ and $-1-2\beta$.
Notice how $\lambda=-4$ is an eigenvalue. This is specific to this graph.
The introduction of the link with coupling $\beta$ is a way to shift
the normal modes of the system. Notice the "crossing" $\lambda_3=\lambda_4$
for $\beta=3/2$. 
The eigenvectors $v_2,~~v_3$ and $v_4$ are shown respectively from 
left to right in Fig. \ref{f1f}. Notice how $v_3$ corresponding to 
$\lambda_3=-1-2\beta$ is localized on the graph while $v_2$ corresponding
to $\lambda_2=-1$ is not localized.

To summarize, we have seen how to generate graphs with soft nodes 
using swivels. From the amplitude equations we inferred that forcing or 
damping will not be effective when applied to these soft nodes.
In the next section we analyze the dynamics close to resonance when
the network is forced or damped at given nodes. We confirm that soft
nodes will not be effective.

\section{Numerical results: forcing the network }

Assume that the graph is forced periodically at a given node $n_f$ and damped 
at a node $n_d$, this happening for a time duration $100 < t< 300$.
This forcing and damping is typical for an electrical
network. The model can also describe an electrical power grid.
The damped nodes correspond to
cities where the energy is absorbed and the forced nodes 
correspond to power stations where energy is introduced in the network. 
We consider a periodic 
forcing for simplicity. The result for other types of forcings 
can be derived from this study using a superposition
argument because the system is linear.

The first interesting observation is that damping or driving 
is ineffective if applied to a soft node. This will happen for
any graph, with symmetries or not. 
The second effect that we will
show is when two eigenfrequencies are
close. This happens for the tree when $\alpha=1$ or for the 
graph with a closed swivel when $\beta\approx 0$ or $\beta=3/2$.
Some degree of symmetry is usually necessary for this.
Then the system can function on two different eigenmodes
when it is forced. We will see that the forcing determines 
which mode is selected.

\begin{figure}
\centerline{
\epsfig{file=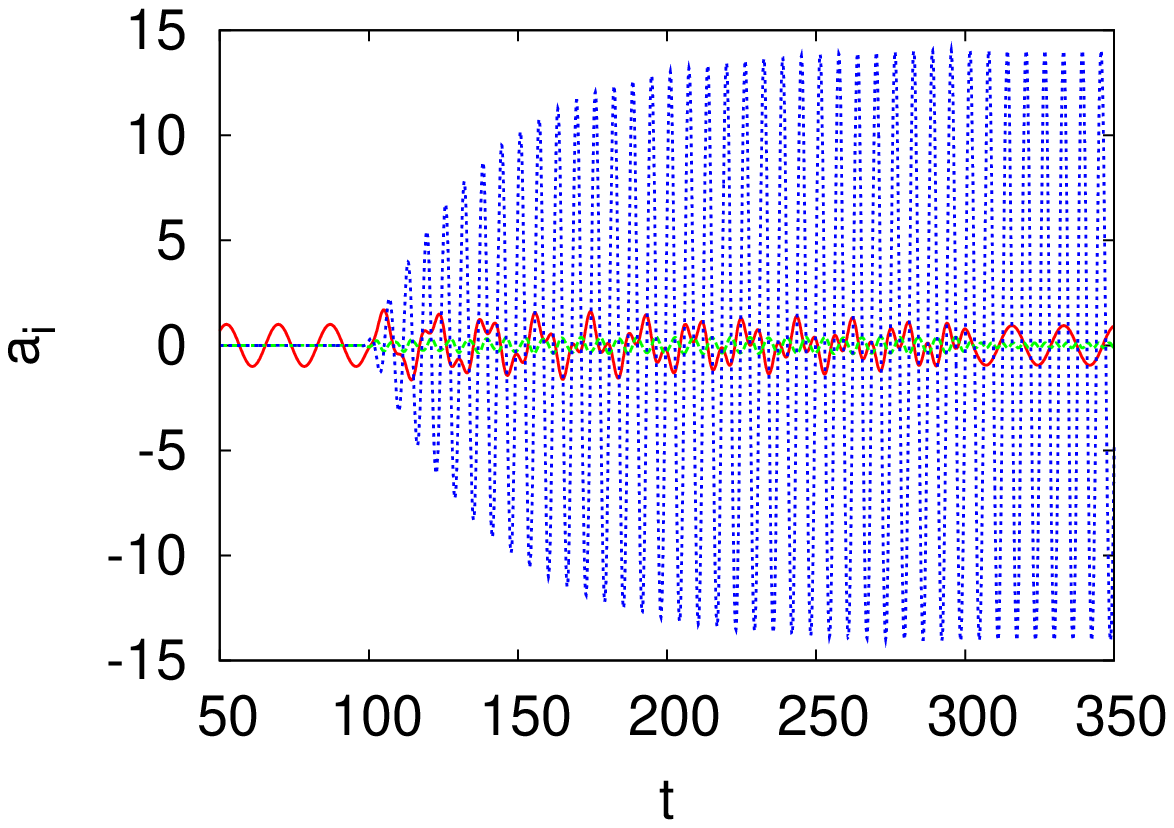,height=5 cm,width=7 cm,angle=0} 
\epsfig{file=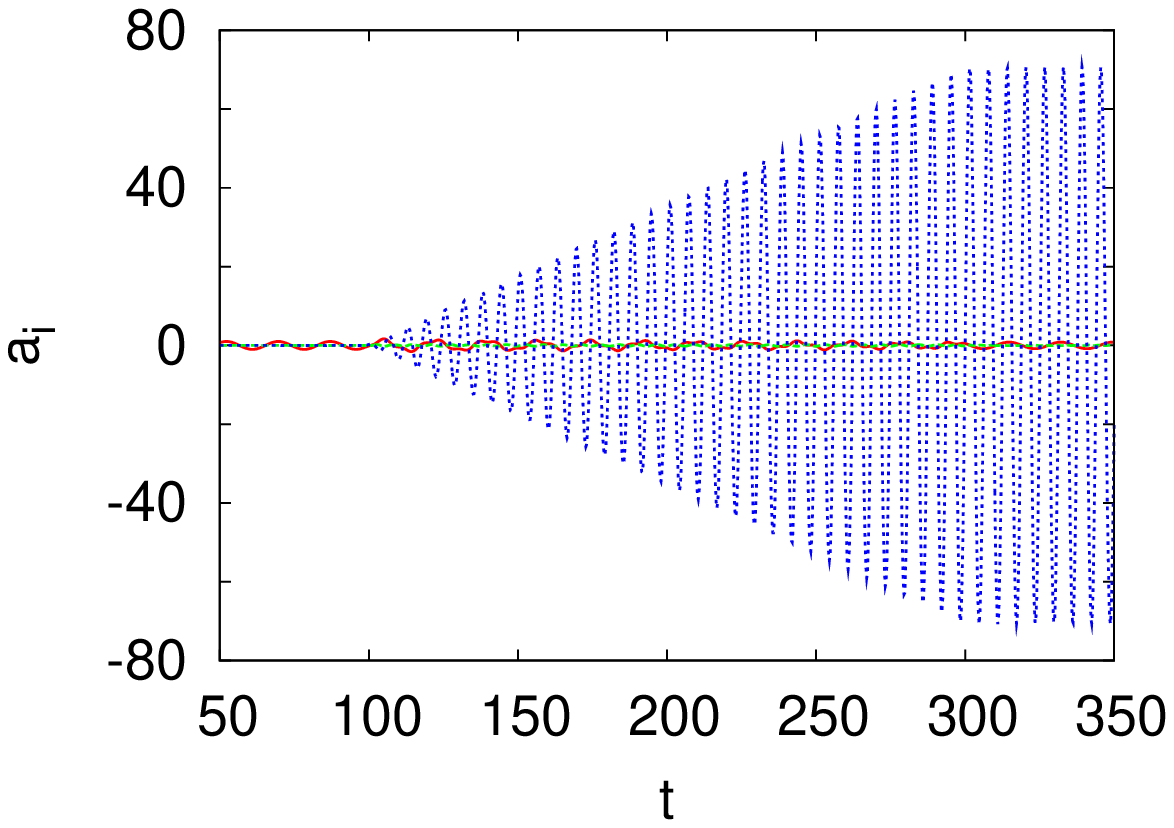,height=5 cm,width=7  cm,angle=0} 
}
\caption{ Example of the influence of damping applied to different nodes,
the graph is a tree with $\alpha =0.1,~\beta=0$.
Plot of $a_2(t)$ (blue online), $a_3(t)$ (red online) and 
$a_4(t)$ (green online) when
the node 4 is excited with a frequency
$\omega = 1$. The left (resp. right) panel corresponds to the node 1
(resp. the node 2) being damped.
}
\label{f4}
\end{figure}

To illustrate the first effect we choose a tree with 
$\alpha=0.1,~\beta=0$.
Fig. \ref{f4} shows the different mode amplitudes $a_i(t)$ 
when exciting the tree
on node 4 and damping it on node 1 (left panel) and damping
it on node 2 (right panel).
For the damping on node 1 the corresponding amplitude equations are
\begin{eqnarray}
{\ddot a_1}=  d( -0.25{\dot a_1}    -0.15{\dot a_3}    
-0.2{\dot a_4}    -0.35{\dot a_2})+    0.5f,\\
{\ddot a_2}+  a_2=d(-0.35{\dot a_1}
-0.21{\dot a_3} -0.29  {\dot a_4} -0.5{\dot a_2} )
   -0.7 f,\\
{\ddot a_3} + 
0.13a_3= d(-0.15{\dot a_1} -0.1 {\dot a_3}
-0.12{\dot a_4}    -0.21{\dot a_2})+ 0.3f,\\
{\ddot a_4} +  3.1 a_4 =d(-0.2{\dot a_1} 
-0.12{\dot a_3} -0.16{\dot a_4} -0.28{\dot a_2})
+0.4    f.
\end{eqnarray}

The second equation is resonant if $f = \sin t$ ($\omega=1$ ) but it is damped. This
gives rise to an increase of the amplitude of mode 2 with a saturation
as shown in the left panel of Fig. \ref{f4}. The same effect happens if
the node 4 is damped. 
The final value of $a_2$  can be obtained
by a Green's function approach.  For that we neglect the other modes
and write the $a_2$ amplitude equation as
\be\label{damped_lin_reson}
{\ddot a_2}+  \omega_0^2 a_2 = f(t) -d {\dot a_2}.\ee
The Green's function $G(t-t_0)$ solves
$${\ddot G} + \omega_0^2 G +d {\dot G} = \delta(t-t_0),$$
where $\delta(t-t_0)$ is the Dirac function at $t_0$. From
\cite{Churchill} we have
\be\label{green}
G(t-t_0)= e^{-{d\over 2}(t-t_0)} 
{\sin \sqrt{\omega_0^2 -{d^2\over 4}}(t-t_0) \over \sqrt{\omega_0^2 -{d^2\over 4}}}.\ee
The solution is then given by
\be\label{sol_green}
a_2(t)= \int_0^t G(t-t_0) f(t_0)dt_0.\ee
This expression agrees with the plot on the left panel of Fig. \ref{f4}.
When damping is applied to nodes 3 or 4, the equation for
$a_2$ has no damping, it is 
\be 
{\ddot a_2}+  a_2= -0.7 f.\ee
The amplitude of the mode 2 then grows linearly
as expected. This is shown in the right panel of Fig. \ref{f4}.
The final amplitude in Fig. \ref{f4} can be easily found by 
seeing that the solution of the linear resonance equation 
\be\label{lin_reson}
{\ddot a_2}+  a_2 = f \sin t,\ee
is 
\be\label{sol_lin_reson}
 a_2 (t) = - {f\over 2} t \cos t .\ee
In the example above one gets 
$$max( a_2(300)) = - {0.7 \over 2} (300-100) = 70,$$
which is in excellent agreement with the value on the right panel of
Fig. \ref{f4}. This unbounded growth leads to the destruction
of the network because damping is ineffective. 
This will also occur for the graph with a closed swivel damped at node 2 when 
$\omega\approx \omega_2$ or $\omega\approx \omega_3$
as shown in Fig. \ref{f1f}.
This enumeration shows that this non effectiveness of the damping (or
driving) can occur for any network, the only ingredient necessary
is the presence of a "soft node".

\begin{figure} [H]
\centerline{\epsfig{file=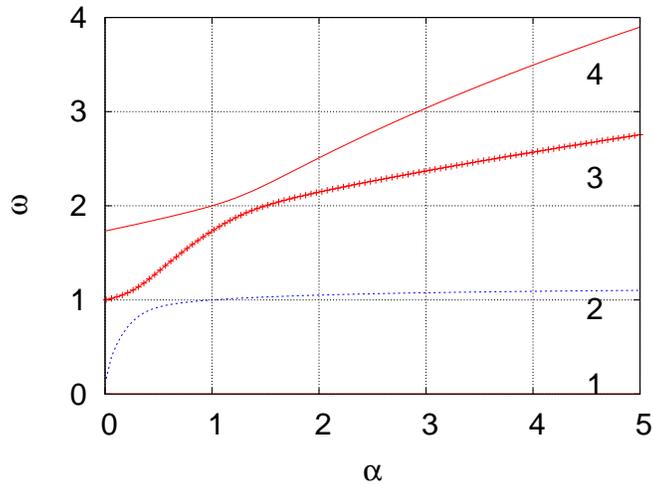,width=0.8\linewidth,angle=0}}
\caption{ Plot of the eigenfrequencies $\omega_i,~i=1-4$ as a function of
$\alpha$ for the graph shown in Fig. 3 with $\alpha=\beta$ i.e. 
a graph with a cycle. 
}
\label{f1d}
\end{figure}

\begin{figure} [H]
\centerline{ \epsfig{file=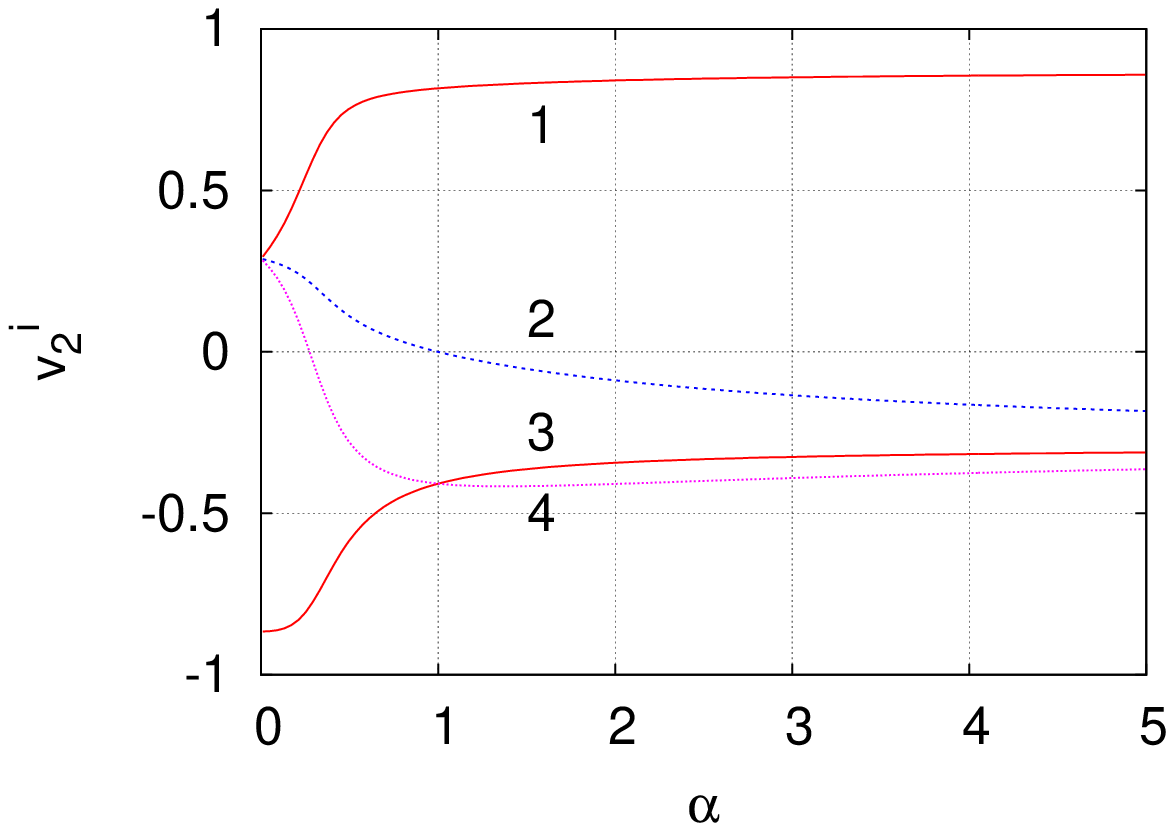,height=4 cm,width=10cm,angle=0}}
\centerline{ \epsfig{file=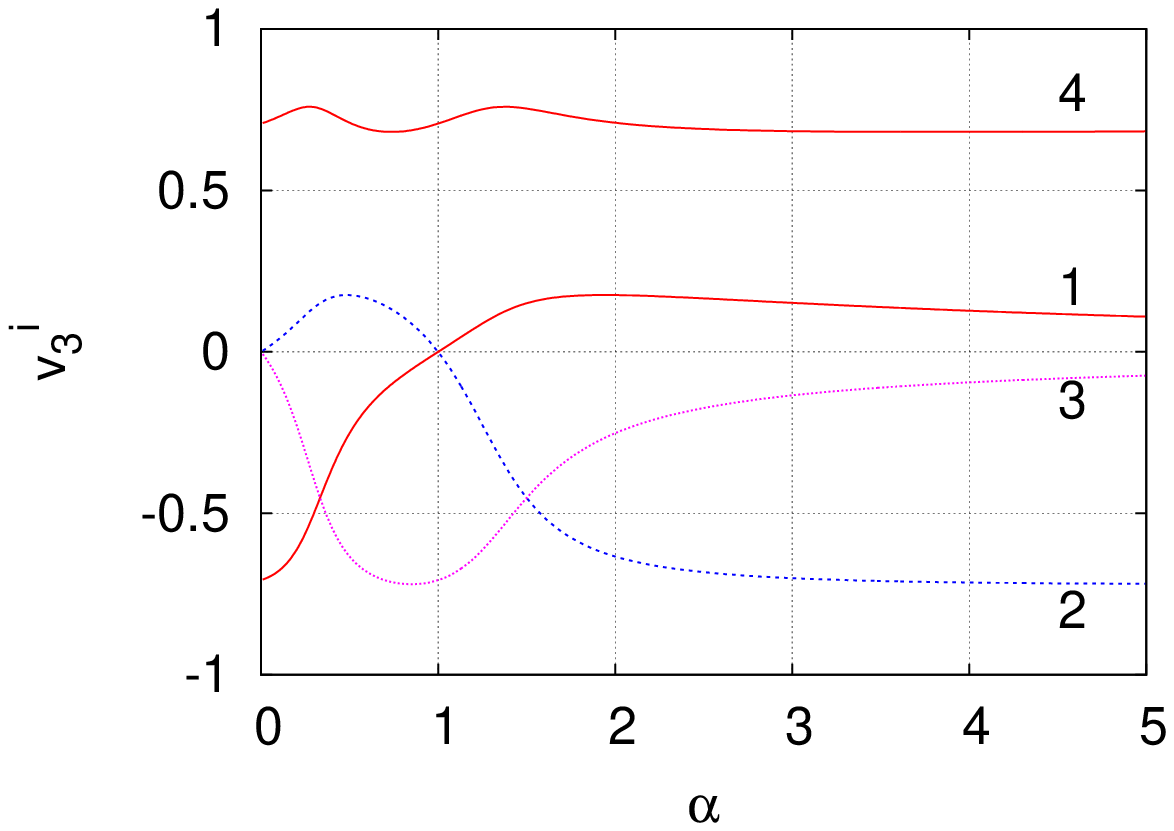,height=4 cm,width=10 cm,angle=0}}
\centerline{ \epsfig{file=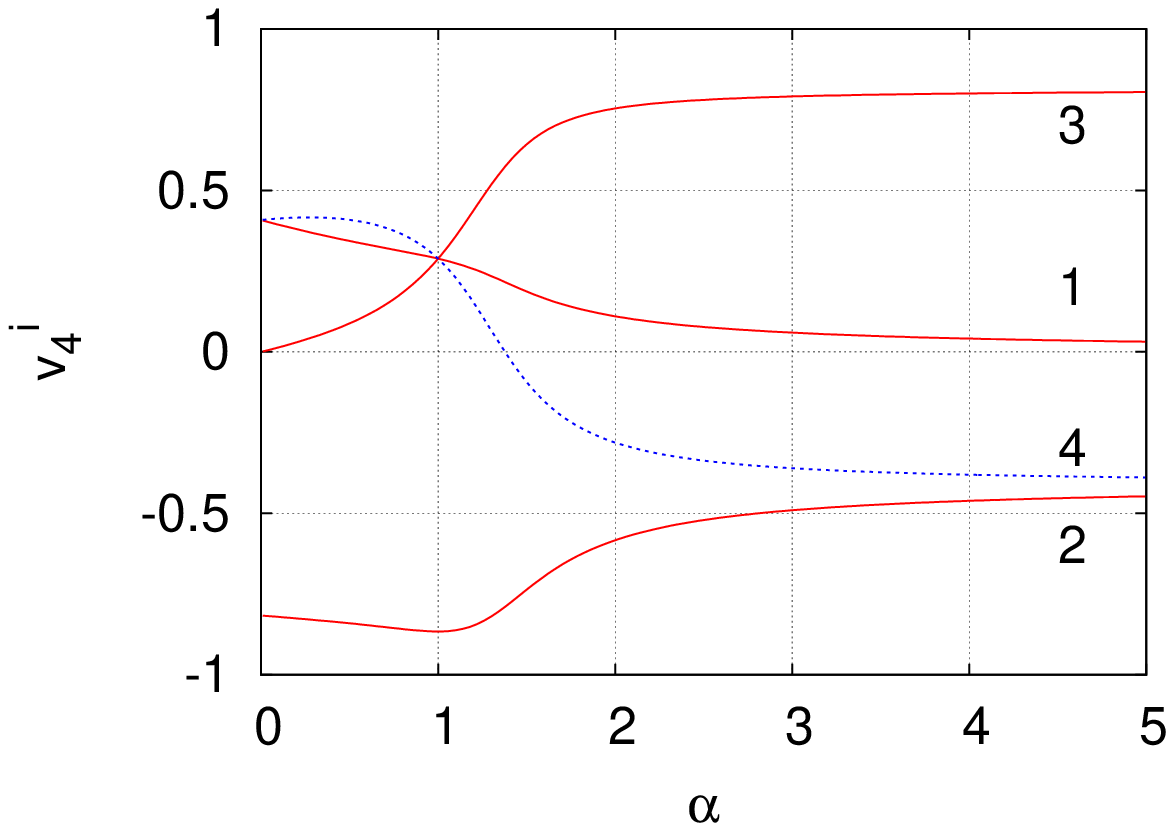,height=4 cm,width=10 cm,angle=0}}
\caption{Plot of the components of the non trivial eigenvectors $v_2,v_3,v_4$ from top to bottom
as a function of $\alpha$. 
}
\label{f2a}
\end{figure}

\begin{figure}[H]
\centerline{ \epsfig{file=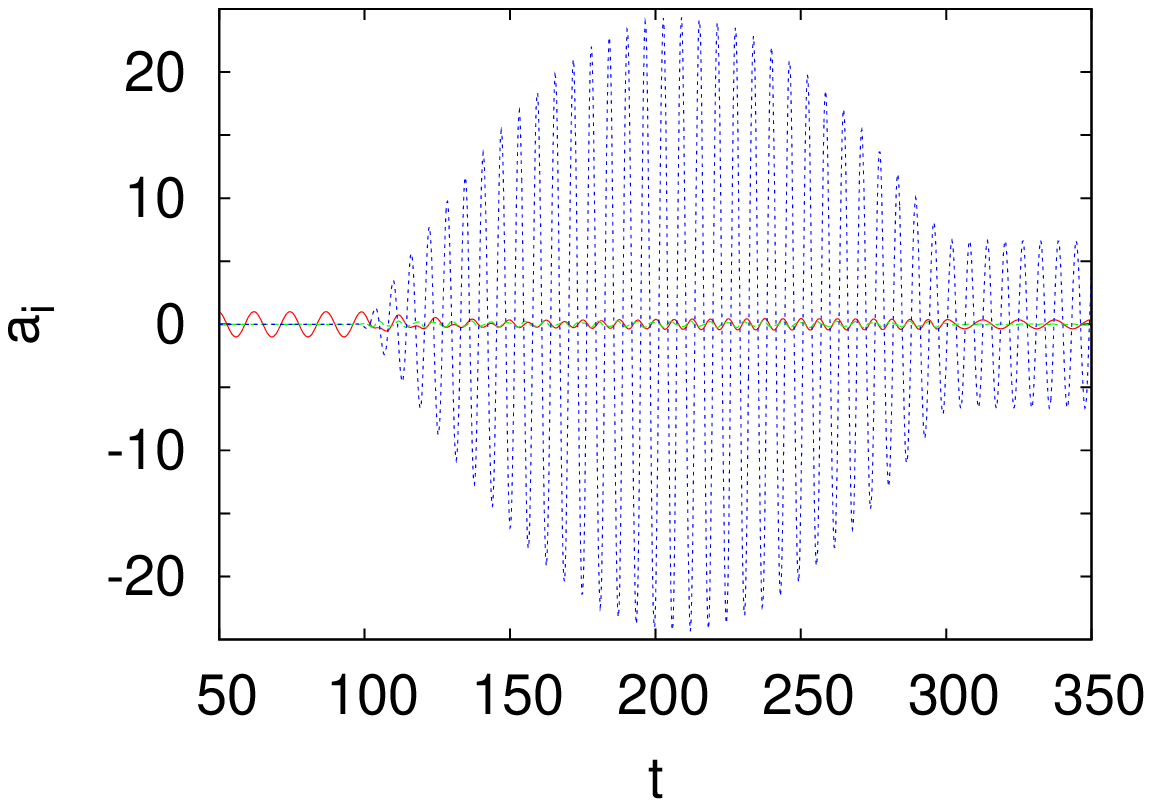,height=6 cm,width=6 cm,angle=0}
 \epsfig{file=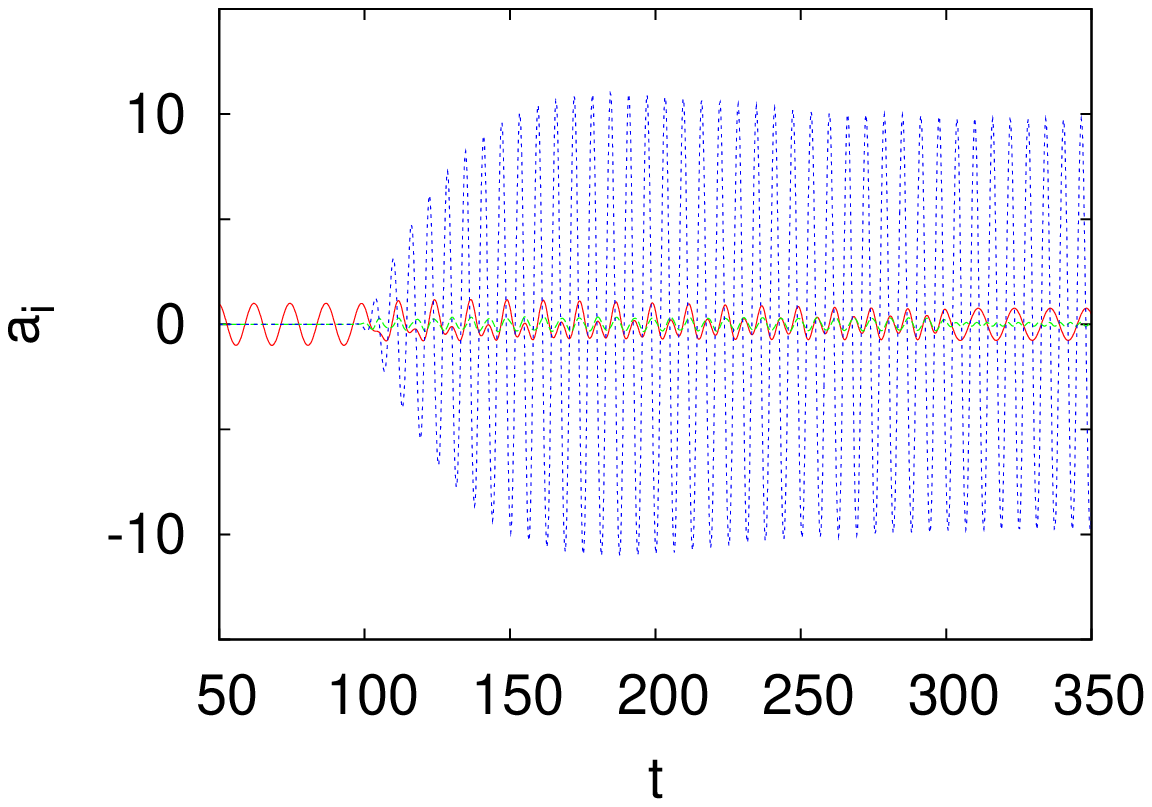,height=6 cm,width=6 cm,angle=0}}
\caption{ Similar plot as in Fig. \ref{f4} except that the graph
now has a cycle $\alpha=0.1,~\beta=0.1$.
Plot of $a_3(t)$ (blue online) and $a_2(t)$ (red online) 
when the node 4 is excited with a frequency
$\omega = 1$. The left (resp. right) panel corresponds to the node 2
(resp. 1) being damped.}
\label{f7}
\end{figure}

Even when the node is close to soft do we get this strong reinforcement
of the oscillations. Consider the graph of Fig. \ref{f1b} with $\beta=\alpha$.
It has no particular symmetries and no exact soft nodes. For this more 
complex graph, the eigenvalues and eigenvectors must be computed 
numerically. We have used Matlab.
The plot of the
eigenvalues as a function of $ \alpha$ is shown in Fig. \ref{f1d}.
As expected from the theory \cite{mohar91}, the eigenvalues of the graph
with a cycle $\mu$ and the eigenvalues of the tree $\lambda$ are
interlaced such that
\be\label{interlace}
0=-\lambda_1=-\mu_1 \le -\lambda_2 \le -\mu_2 \le -\lambda_3 \le -\mu_3
\le -\lambda_4 \le -\mu_4.\ee
Note that $\lambda_4 = \mu_4$ for $\alpha <1$. The fact that $\lambda_2$ and
$\lambda_3$ are close confines $\mu_2$. 
The dependency of the eigenvectors on the coupling parameter $\alpha$
is shown in Fig. \ref{f2a}. 
As shown in the middle panel of Fig. \ref{f2a} for $\alpha=0.1 =(\beta)$
and mode 3 node 2 is close to soft while node 1 is non zero. We force the
network at frequency $\omega=1 \approx \omega_3$ to be close to resonance.
When the graph has a cycle so that $\alpha=\beta=0.1$ we get a slightly
different picture because the linear resonance observed previously
is not exact. There is a small damping due to the non zero
second coordinate of the eigenvector $v_3$.
We excite node 4 at a frequency $\omega=1$ and damp the
nodes 1,2,3 and 4 respectively. We find two groups of behaviors
shown in Fig. \ref{f7},  the left panel corresponds to
damping on nodes 2 and 3. As expected the maximum amplitude of mode
$a_3$ is much larger than when nodes 1 and 4 are damped (right panel). This is
because nodes 2 and 3 are almost soft and not nodes 1 and 4.
One can see the typical beat at
frequency $(\omega_3-\omega)/2$ which yields a half-period
$T/2 = 157$. The mode 2 which is the initial condition
is strongly reduced.
Note that on the left panel the amplitude of mode 3 is
practically constant during and after the forcing/damping region.
\noindent This shows that approximate soft nodes exist also in graphs without
a particular symmetry. Fig. \ref{f2a} shows that for large $\alpha$ nodes
1 and 3 are almost soft for $v_3$ and node 1 is almost soft for $v_4$.

\begin{figure}[H]
\centerline{ \epsfig{file=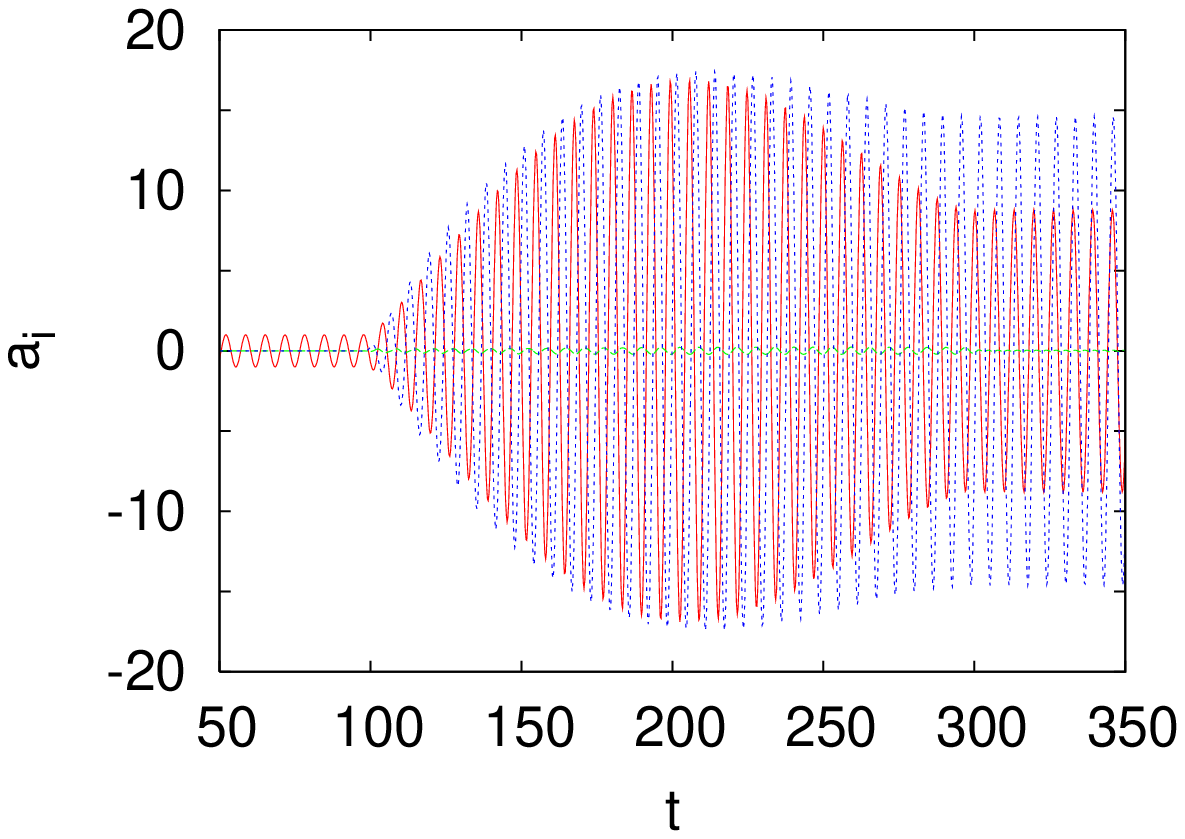,height=6 cm,width=4 cm,angle=0}
\epsfig{file=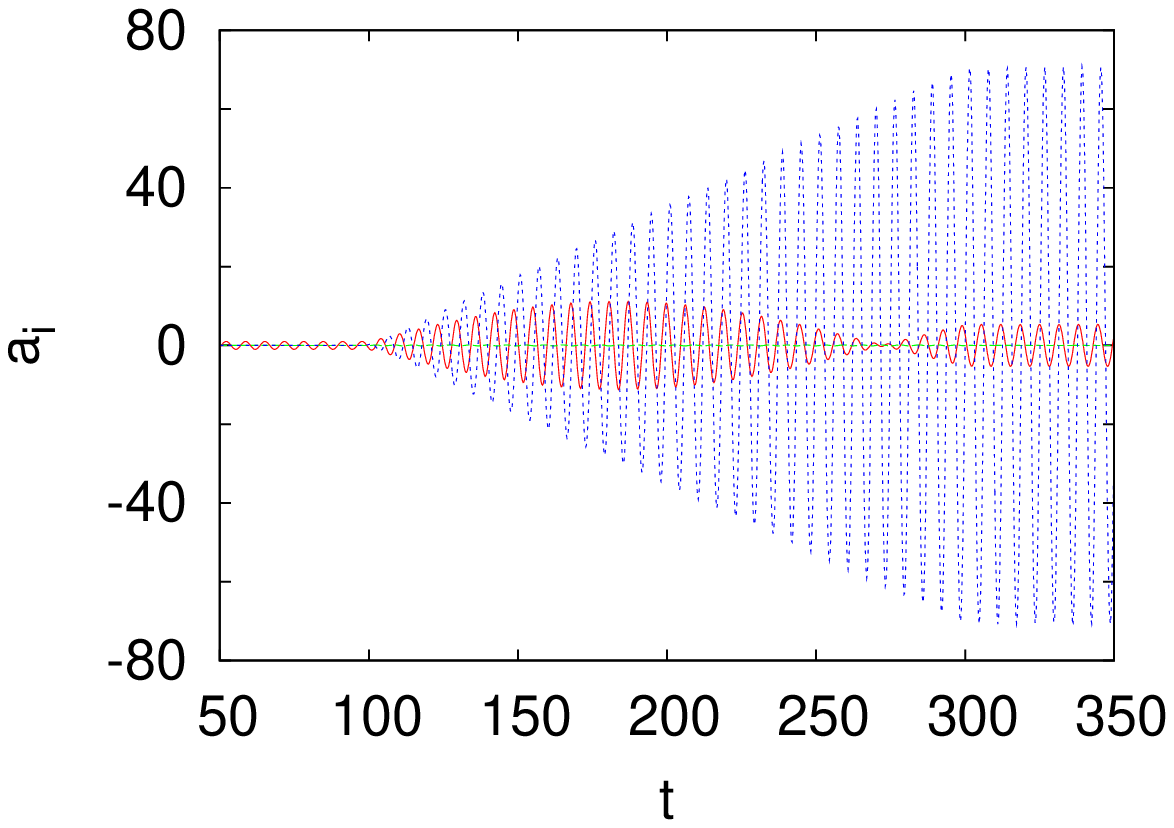,height=6 cm,width=4 cm,angle=0}
\epsfig{file=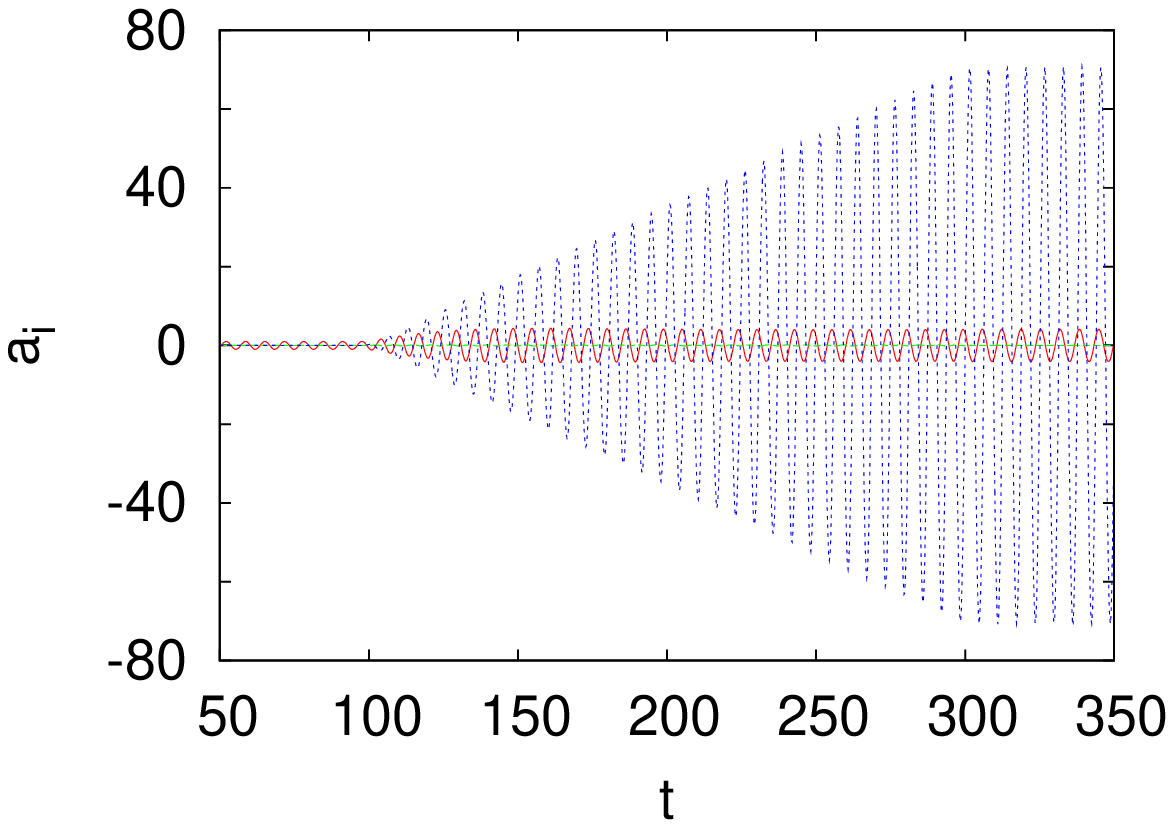,height=6 cm,width=4 cm,angle=0}}
\caption{ 
Effect of a multiple eigenvalue in the Laplacian $G$.
Plot of $a_2(t)$ (blue online) and $a_3(t)$ (red online)
when the node 4 is excited with a frequency
$\omega = 1$. The panels correspond respectively to damping the 1st node 
(left), damping the second node (middle) and damping the 3rd node (right).
The parameters are $\alpha=0.9,~\beta=0$ and the system is started
on the third mode $v_3$.}
\label{f9}
\end{figure}

Another effect that occurs in fairly symmetric graphs is that eigenfrequencies
can "cross" as a parameter is varied. This means that two close frequencies
can correspond to two eigenmodes that can be very different. If one
of the modes has a soft node and the other does not, one can control which
mode will appear by selectively exciting or damping a given node of the
network. As an example consider the tree graph
with $\alpha=0.9, ~\beta=0 $. As shown in Fig. \ref{f1c} we have
two close eigenvalues 
$$\omega_2=1, ~ v_2= \left ( \begin{matrix}
0.71 \cr
0 \cr
0 \cr
-0.71
\end{matrix}  \right)
~~, \omega_3=0.964, ~ v_3= \left ( \begin{matrix}
0.4 \cr
0.028 \cr
-0.82 \cr
0.4
\end{matrix}  \right).
$$
Mode 2 has $s=2$ and 3 as soft nodes while mode 3 has $s=2$ almost soft.
We have excited node 4 with a frequency $\omega=1=\omega_2\approx \omega_3$.
The time evolution of the modes $a_2$ and $a_3$ are shown in Fig. \ref{f9}
respectively in light and dark grey (blue and red online). 
When node 1 is damped, both modes $v_2$ and $v_3$ can be excited because
they have non zero components on that node. This is shown in the left panel
of Fig. \ref{f9}. On the contrary when node 2 is damped (middle panel of 
Fig. \ref{f9}) no damping occurs for the amplitude $a_2$ causing an
unbounded linear growth. The mode $v_3$ is excited but in a much smaller way
because it is weakly damped, since $v_3$ has a small non zero component
on node 2. When node 3 is damped (right panel of Fig. \ref{f9}),  the 
mode 3 is strongly damped so that
there is practically only mode 2. Therefore one sees that applying damping
to node 3 will result in mode 2 only being excited while applying
damping on node 1 will result in the presence of both modes 2 and 3.

\section{Conclusion }

To describe the flow of a miscible quantity on a network, we considered the graph wave
equation where the standard continuous Laplacian is replaced by the graph Laplacian.
We have only considered a node wave equation. There would be a similar branch wave
equation.
We showed that a natural example is an electrical network of inductances on the branches 
and capacities on the nodes. This system can also describe shallow water waves on a network 
of canals or fluid flow in a network of pipes. There is also a mechanical analog in terms
of masses and springs. In general this graph wave equation describes the small
oscillations of a network around it's functioning point.

Since the graph Laplacian is a symmetric matrix, its eigenvectors can be chosen to be 
orthogonal. These provide
a natural basis to describe the flow in terms of amplitudes on each mode. We derived such amplitude
equations when the network is forced and damped on a given node.
The eigenvalues and components of the eigenvectors are important elements of these amplitude
equations. The amplitude equations revealed the importance of soft
nodes where one of the eigenvector components is zero. Any action like forcing
or damping on such a node is ineffective. The concept was formalized.
We also showed that a sufficient condition for a graph to have
a soft node is to have a swivel. Approximate soft nodes can also
occur when the couplings depend on a parameter.

The numerical analysis of the amplitude equations when the network is
forced at resonance confirmed the importance of
soft nodes. In particular if damping is applied to such a node, the 
network will go into an unbounded resonance which has catastrophic effects.
The network can also have multiple eigenvalues. If one of the eigenmodes has a soft node, we showed how the system can be controlled to have one
mode only appear or to have two modes appear.

These soft nodes will appear in graphs with symmetries, for example swivels.
They can also be found in graphs where the couplings depend on a parameter.
This study can then be applied to complex physical networks, like
a power grid.

\vskip .5 cm

{\bf Acknowledgements}

The authors thank the Region Haute-Normandie for support through
the grant "Statique et dynamique de r\'eseaux simples" (GRR-TLTI). Elie Simo
thanks the Laboratoire de Mathematiques de l'INSA for its hospitality
during two visits in 2008 and 2009. The authors thank the Centre de 
Ressources Informatiques de Haute-Normandie for the use of their
computing ressources.

\end{document}